# Plasmonic Structures, Devices, and Integrated Applications: A Patent Analysis

**Mahdi Javidnasab[1], Sajjad Hosseinzade[1]**

1. Department of Condensed Matter Physics, Faculty of Physics, University of Tabriz, Tabriz 51666-16471, Iran

**Abstract**

This review systematically analyzes patent disclosures regarding plasmonic structures, devices, and integrated applications, highlighting the technology's capability to confine and manipulate electromagnetic energy at the nanoscale. Core materials rely on highly conductive noble metals (Gold, Silver), alongside innovative alternatives like Transparent Conducting Oxides (TCOs), Graphene, and 2D materials. These materials are engineered into diverse nanostructures that exploit Surface Plasmon Resonance (SPR) and Localized Surface Plasmon Resonance (LSPR). LSPR generates intense electromagnetic "hot spots" necessary for signal amplification (SERS) and provides extreme refractive index sensitivity for label-free detection.

Key device elements include subwavelength plasmonic waveguides (hybrid and slot types) for optical signal routing. Dynamic control is achieved through active tuning mechanisms, utilizing electrical gating of charge carrier density (e.g., TCOs, Graphene) and solid-state phase change materials (e.g., VO2).

**The patent applications span critical technological sectors:**

1. Biosensing and Diagnostics: Enabling label-free detection of biomarkers (e.g., cancer markers, nucleic acids, pathogens) through resonance shifts, and achieving amplification via SERS and MEBL, often integrated into portable point-of-care (POCT) platforms.
2. Data Storage and Communication: Increasing recording density in Heat-Assisted Magnetic Recording (HAMR) via Near-Field Transducers (NFTs) that focus light to spots <10 nm. Plasmonic waveguides serve as interconnects in Electronic-Photonic Integrated Circuits (EPICs), supporting ultra-fast modulation and THz generation.
3. Energy and Optics: Enhancing efficiency in Photovoltaics (light trapping and hot carrier generation) and Thermophotovoltaics (spectral control). Devices also include deep subwavelength nanolasers and enhanced solid-state lighting (OLEDs).
4. Biomedical Interventions: Utilizing photothermal effects for targeted cancer treatment (Plasmonics Enhanced Phototherapy) and employing non-invasive systems for real-time monitoring of living excitable cell activity (e.g., cardiac cells).

The scope of these patent disclosures confirms plasmonics' essential role in advancing highly integrated, sensitive, and ultrafast technologies

# Contents

# 1 Plasmonic Materials & Components

## 1.1 Plasmonic Materials

Plasmonic materials, characterized by their ability to exhibit surface plasmon resonance (SPR) when excited by electromagnetic energy, enable the confinement and manipulation of light at the nanoscale. These materials typically possess free carriers, such as free electrons, holes, or conduction band electrons, and include metals, semiconductors, molecules, insulators, or single-atom species in specific configurations. This document categorizes plasmonic materials into four primary groups: (1) metals and alloys, (2) alternative plasmonic materials, (3) two-dimensional (2D) and quantum materials, and (4) structural and functional components.[1], [2], [3], [4], [5], [6], [7], [8], [9], [10], [11], [12], [13], [14]

### 1.1.1 Metals and Alloys

Metals are the predominant traditional materials for exploiting plasmonics.

**Noble Metals:**

The most commonly cited plasmonic metals are those known for their high conductivity and plasmon supporting capabilities:

- **Gold (Au):** Frequently listed as a suitable plasmonic material. Gold nanorods or gold antennas are exemplary plasmonic nanomaterials used to form arrays. Gold nanospheres are also mentioned in plasmonic laser nanoablation studies.[1], [4], [5], [7], [13], [15], [16], [17], [18], [19], [20], [21], [22], [23], [24], [25], [26], [27], [28], [29], [30], [31], [32]

- **Silver (Ag):** Highly favored, particularly because it can produce a narrow plasmon resonance, making it advantageous for switching applications (e.g., between digital one and zero states). Silver is also widely cited as a primary plasmonic material. Silver nanostructures are used on graphene via thermally assisted self-assembly. [1], [4], [5], [6], [7], [9], [13], [15], [16], [17], [18], [19], [20], [21], [22], [23], [24], [25], [28], [29], [33], [34], [35]

- **Copper (Cu):** Listed as a plasmonic metal. A copper plasmonic material may feature a copper oxide shell ($CuO$ *or* $Cu_{20}$) surrounding a copper core.[1], [5], [7], [13], [15], [16], [17], [18], [19], [20], [21], [22], [23], [24], [25], [29], [35]

- **Platinum (Pt) and Palladium (Pd):** Also identified as plasmonic metals.[1], [7], [15], [21]

**Other Plasmon-Supporting Metals and Alloys:**

- **Aluminum (Al):** A plasmon-supporting metal. However, in the Terahertz (THz) range, aluminum can exhibit "unreasonably high propagation losses". Aluminum oxide $Al_2O_3$ is used as a buffer layer or waveguide material in near-field light generating devices.[1], [7], [4], [10], [15], [16], [17], [19], [20], [21], [24], [25], [35]

- **Refractory/High-Melting Point Metals**: Ruthenium (Ru), Rhodium (Rh), Iridium (Ir), and Tantalum (Ta) are mentioned as plasmonic materials or as components for molds and diffusion barriers in NFT fabrication. [1], [13], [16], [20], [21], [36]

- **Liquid Metals**: For tunable plasmonic arrays, a liquid metal, such as a eutectic alloy composed of Gallium (Ga), Indium (In), and Tin (Sn), can be used. Liquid metals offer the advantage of being tunable after fabrication, unlike fixed solid metals. [10]

**Alloys:** Plasmonic materials often include alloys, which can be binary ($e.g., AuAg, AuPd, AgPd, AuCu, AgCu$), ternary, or quaternary. A Rhodum-Aluminum (Rh − Al) metallic alloy is noted for providing a broadband resonance response, which may be suitable for producing an analog output. .[5], [6]

## 1.1.2 Alternative and Doped Plasmonic Materials

Beyond metals, alternative materials such as doped semiconductors, transparent conducting oxides (TCOs), metal oxides, and nitrides expand the plasmonic material landscape, offering stability and spectral versatility.[7], [8]

**Transparent Conducting Oxides (TCOs) and Metal Oxides/Nitrides:**

These materials function as plasmonic materials, often doped, and are sometimes preferred over noble metals for stability or spectral range:[7], [8]

- **Indium Tin Oxide (ITO):** Mentioned as a plasmonic material and as a doped semiconductor. It is one example of a transparent conducting oxide.[1], [5], [7], [8], [20]

- **Aluminum Zinc Oxide (AZO):** Mentioned as an exemplary plasmonic nanomaterial, typically used as nanoparticles. It is also listed as a transparent conducting oxide/doped semiconductor.[1], [7], [8], [30], [31]

- **Other Doped Oxides:** Fluorine-doped Tin Oxide (FTO), Indium-doped Cadmium Oxide (ICO), Antimony-doped Tin Oxide (ATO), and Cesium Tungsten Oxide ($Cs_xWO_3$) are all identified as plasmonic oxide materials or TCOs.[5], [7]

- **Metal Oxides:** Materials like Indium oxides ($In_2O_3$) , Tin oxides ($SnO_2$), Titanium oxides($TiO_2$), $Zirconium oxides$($ZrO_2$), $Cesium oxides$ ($CeO_2$), $Zinc oxides$ ($ZnO$), Copper oxides ($CuO$), $and Gallium oxides$ $Ga_2O$ can be used to make plasmonic nanomaterials, often being resistant to high temperatures.[31]

- **Metal Nitrides:** Titanium Nitride (TiN) and Zirconium Nitride (ZrN) are examples of plasmonic materials.[1], [5]

- **Dopants:** Metal oxides or sulfides can be doped with elements such as Pd, Pt, Au, Sn, Al, Nb, or Ta.[30], [31]

### 1.1.3   Nanostructures and Quantum Materials

Plasmonic components often rely on the precise geometry and interaction between specific material combinations at the nanoscale.

**Nanostructures and Nanoparticles:** Plasmonic nanoparticles can be synthesized in various shapes—including nanospheres, nanorods, nanocubes, nanopyramids, nanoshells, and nanostar particles.[21], [37]

- **Core/Shell Structures**: Plasmonic nanomaterials can be nanoparticles having a single material core structure or an inner core and an outer shell comprising the same or different materials. Silver wire cores surrounded by a $SiO_2$ shell and an outer shell of "$\{CdSe\}$ nanocrystals form a composite metal-insulator-semiconductor nanowire system used as a waveguide for 1D-surface plasmons.[30], [31], [38]

- **Nanowires/Nanoparticles:** Plasmonic components may be composed of metallic nanostructures (e.g., silver nanoislands) or nanowires (e.g., Ag, Au, or Cu nanowires acting as waveguides). Noble metal bipyramid nanocrystals are used as plasmonic heaters for photothermal radiation-to-heat conversion.[9], [11], [39]

**Quantum and 2D Materials:**

- **Quantum Dots (QDs):** Colloidal nanocrystals like CdSe/ZnS Quantum dots (QDs) show metal-enhanced fluorescence when resonantly coupled to surface plasmons. QDs are key in devices involving plasmonic coupling for optical emission/enhancement.[11], [18], [38], [40]

- **Transition Metal Dichalcogenides (TMDs):** Two-dimensional materials like monolayer $WS_2$, $MoS_2$, and $WSe_2$ are vital for electronic-photonic integrated circuits (EPICs). TMDs are suitable because they can emit and absorb light at the same wavelength via direct excitonic transition.[9], [41]

- **Graphene**: Recognized as a 2D plasmonic material. Graphene plasmonics can yield huge optical enhancements and are central to components like graphene-based plasmonic slot electro-optical modulators.[5], [7], [10], [25]

- **Perovskites, Silicides, Germanides**: Also listed as examples of plasmonic materials.[7]

## 1.1.4 Materials for Control, Integration, and Structure

**Waveguide and Interface Materials:** Plasmonic materials are often integrated with dielectric or semiconductor materials for functional devices:

- **Tantalum Pentoxide ($Ta_2O_5$) and Aluminum Oxide ($Al_2O_3$):** Used as waveguide and buffer layer materials in near-field light generating devices (NFTs), particularly in heat-assisted magnetic recording (HAMR). An adhesion layer, such as titanium, may be placed between a buffer layer like $Al_2O_3$ and a noble metal plasmonic layer (e.g., Ag or Au) to prevent exfoliation.[4]

- **Insulators/Dielectrics:** The confinement of surface plasmon waves occurs at a metal-dielectric interface. Silicon dioxide ($SiO_2$) is a common material, sometimes forming a shell around metal cores.[2], [23], [38]

**Modulation and Switching Materials:**

- **Solid-State Phase Change Materials:** Materials like $VO_2$ ($Vanadium\ Dioxide$) or GeSb films are incorporated into plasmonic modulators to actively modulate light propagation by changing phase.[12]

- **Photochromic Molecules**: These molecules (e.g., spiropyran-based) are dispersed within a matrix material (like polymethyl methacrylate, PMMA) in nanostructured photonic materials. These undergo reversible isomerization upon light illumination, enabling photoswitchable effects like Rabi splitting for optically rewritable photonic waveguides.[14]

**Structural Components:**

- **Molds/Barriers**: Materials for fabrication structures, such as diffusion barriers (rhodium, tungsten, tantalum, tantalum nitride, ruthenium, titanium, titanium nitride) or molds used in forming NFTs (W, Ta, Ti, Pt, Pd, Ru, Ir, C, oxides, nitrides, carbides, or borides) are critical components.[13], [36]

- **Substrates**: Plasmonic components can be formed on various substrates, including glass, semiconductor materials (e.g., silicon), and even paper substrates (cellulose, nitrocellulose).[6], [27], [33]

In summary, plasmonic components rely heavily on metals (Au, Ag, Cu, Al) structured at the nanoscale to control plasmons, but the field is rapidly expanding to include engineered materials like TCOs (ITO, AZO), 2D semiconductors (TMDs, graphene), and complex composite structures for enhanced functionality, tunability, and integration into compact circuits and sensors.[5], [8], [9], [30]

# 1.2 Plasmonic Nanostructures: Geometry, Dimensions, and Functional Applications

Plasmonic nanostructures are critical for exploiting surface plasmon resonance (SPR) or localized surface plasmon resonance (LSPR), enabling light manipulation at subwavelength scales. These nanostructures, defined by their nanoscale dimensions and specific geometries, underpin applications in sensing, waveguiding, and energy manipulation. This document details the geometries, dimensions, and functionalities of plasmonic nanostructures as described in the provided sources.

## 1.2.1 Defining and Characterizing Plasmonic Nanostructures

Definition and Scale: Plasmonic components necessarily rely on nanoscale features due to the nature of plasmon excitation. A "nanostructured" component is defined as any structure with one or more nanosized features, where a nanosized feature has at least one dimension less than 1 $\mu m$ in size. This includes nanowires, nanotubes, nanoparticles, and nanopores, or combinations thereof.[7], [14]

**Dimensions and Size Ranges:** The functional size of these elements is critical as it tunes the plasmon resonance.[21]

- **Nanoparticles (General):** Particles generally range from 1 nm to 100 nm in size. Plasmonic nanoparticles (NPs) can have an average size ranging from 10 nm to 500 nm or even 1 nm to 1000\ nm. For solar-to-chemical efficiency, NPs sized 1 nm to 1000 nm have significantly higher efficiency than micron-sized catalysts ($100 \, \mu m$).[16], [42], [43], [44]

- **Nanocrystals:** Core sizes can range from 5 nm to 100 nm.[30], [31]

- **Nanostructures on Graphene:** For maximizing the plasmonic effect and minimizing Ohmic loss in graphene-based devices, nanostructures typically have a lateral dimension of about 1 nm to 200 nm, with preferred dimensions in the tens to about a hundred nanometers. For silver nanostructures, the optimal lateral dimension is about 50 to 100 nm in a disk or hemisphere shape. The average height is typically 1 nm to 100 nm, with 60 nm being preferred in one example.[22]

- **Nanowires/Nanorods:** Nanorods in arrays can range from 10 nm to 300 nm wide, 20 nm to 2000 nm long, and $10\,m\;to\;500\,m$ thick.[30], [31]
- **Plasmonic Cavities:** Plasmon lasers designed for pure plasmon modes may have extremely scaled gain regions ($thickness\;\sim 2-20\,m$). Sub-wavelength features are often crucial for devices like the metal-dielectric-metal plasmon cavity that squeeze visible light waves into a 3-nm-thick and 55-nm-long space.[41], [45], [46]

## 1.2.2 Geometries and Shapes

Plasmonic nanostructures exhibit a vast diversity of shapes, as their optical and electrical properties are dependent on their specific geometries.[44], [47]

**Particles and Zero-Dimensional Structures:**

- **Nanoparticles (NPs) / Nanocrystals:** These are widely used and can be discrete units. Their shape can be isotropic (e.g., spherical) or anisotropic (e.g., non-spherical).[15], [21], [31], [42], [43], [48]

- **Spheres/Ovoids:** Spheres are common.[15], [21], [42], [48]

- **Rods/Cylinders:** Nanorods,nanocylinders, or rod-like shapes are used, often characterized by high aspect ratio for tuning.[17], [21], [27], [30], [31], [33], [42], [44], [48]

- **Cubes/Polyhedrons:** Nanocubes, polyhedrons, and nano-octahedra.[33], [38], [43]

- **Stars/Pyramids:** Nanostars, nanopyramids, or triangular prisms.[14], [21], [33], [48], [49], [50]

- **Triangles/Polygons:** Nanotriangles and regular polygons.[15], [42], [51]

**Complex and Hybrid Structures:**

- **Core/Shell Structures:** These composites are common, featuring a core and an outer shell made of the same or different materials.[30], [31]

◦ **Nanoshells/Semishells:** Examples include solid nanoshells, hollow nanoshells, semishells, or nanobowls, often composed of a dielectric core (e.g., silica) coated with an ultrathin metallic layer (e.g., gold). The plasmon resonances of spheroidal shells are tunable based on shell thickness and aspect ratio.[21], [37], [38], [44], [48], [52]

◦ **Core-Shell Heterostructures:** Metal (e.g., noble metal) nanoparticle cores can be coated with a Transition Metal Dichalcogenide (TMD) shell (e.g., $Au MoS2$).[30], [31]

• **Holes/Apertures:** Plasmonic nanostructures often involve patterned openings in metal films.

◦ **Nanopores/Nanoholes:** These are nanosized features used in various applications. Arrays of holes permeated into a plasmonic layer form crucial components of certain metasurfaces and layers.[7], [53]

◦ **Nanogaps:** The gap between metallic nanostructures is where intense electromagnetic "hot spots" occur. Nanogaps, including sub-10 nm gaps, are critical for strong light-matter interactions and sensing.[23], [42], [54], [55], [56]

**One-Dimensional and Two-Dimensional Structures:**

• Nanowires (NWs): Used as nanosized features and sometimes forming arrays. Nanowires or nanowire arrays can function as waveguides for 1D-surface plasmons. A planar nanowire network comprises nanowire segments which can have square, circular, elliptical, or complex cross sections.[7], [14], [30], [31], [37], [38], [48], [52], [57], [58], [59]

• **Nanotubes:** Used as nanosized features, including plasmonic-magnetic bifunctional nanotubes.[7], [14], [60]

• **Thin Films/Layers:** Plasmonic devices can utilize thin metal films permeated by spaced-apart holes.[7], [61]

## 1.2.3  Arrays and Periodic Structures

The arrangement of nanostructures, particularly in arrays, is fundamental to many applications, controlling plasmon propagation and interaction.[30], [31]

**Arrays and Unit Cells:**

- **Arrays of Nanoparticles/Antennas:** Plasmonic components frequently use arrays of nanostructures, such as gold nanorods or gold antennas, nanowire arrays, or plasmonic nanoparticles.[30], [31], [42], [62]

- **Plasmonic Crystals/Metamaterials:** These are periodic geometric features designed to manipulate light, often comprising metallic fingers, dots, crosses, wires, or conductors arranged on a semiconductor surface. Plasmonic crystals can exhibit Tamm states and plasma oscillations in a periodic two-dimensional electron gas (2DEG).[63], [64]

- **Unit Cell Definition:** An array of spaced-apart plasmonic particles is defined by a repeating unit cell, which is the smallest group constituting the pattern. The unit cell can be in the shape of a rectangle. The size of the unit cell defines the distance separating neighboring particles.[14]

- **Arrays of Holes:** Layers permeated by holes often form arrays (e.g., in nanostructured plasmonic materials or metadevices).[7], [61]

**Functionality of Arrays:**

- **Electro-optical devices:** Arrays are used in applications ranging from directional emitters/absorbers to ultra-smooth patterned metal surfaces for metrology.[2]

- **Light Confinement and Focusing:** Arrays of plasmonic nanoantennas can couple propagating waves and surface waves. They can also be collectively heated to create microfluidic vortices for trapping and sorting particles (e.g., nanoantenna arrays on a substrate).[61], [62]

- **Lensing and Focusing:** The design of plasmonic nano-antennas in arrays can be used to build lenses. Plasmonic lenses can be patterned structures that concentrate incident optical energy using surface plasmon polaritons.[61], [65]

## 1.2.4 Nanofeatures for Specific Functions

**Near-Field Transducers (NFTs):** NFTs are critical components in applications like heat-assisted magnetic recording (HAMR). These structures are themselves nanoscale and typically include plasmonic material (e.g., $a\ peg$).[16]

- Internal Nanoparticles: NFTs can incorporate nanoparticles ($1 - 100\ m$) within the plasmonic material (e.g., gold containing ZrO nanoparticles).[16]

- **Nanoparticle Placement:** Nanoparticles may be preferentially disposed on the crystal grain boundaries of the crystalline plasmonic material to improve mechanical strength or block diffusion, sometimes only in a portion of the NFT (e.g., the top or bottom of the peg).[16]

**Waveguides and Light Guiding:** Plasmonics fundamentally allows for light guiding below the diffraction limit.[2], [8]

- **Waveguide Structures:** Plasmonic waveguides are realized through various nanostructures, such as the plasmonic stripe, wedge, slot, or nanowire waveguides.[8]

- **Hybrid Waveguides:** Structures combining plasmonic materials with dielectric waveguides (e.g., hybrid plasmon-waveguide modes) offer low optical loss and subwavelength confinement.[14]

- **Optical Fiber Integration:** Plasmonic nanocircuits, including complex structures like gap plasmonic waveguides, multi-channel plasmonic waveguides, and resonant guided wave networks, can be formed directly on the facet of an optical fiber core using nano-fabrication techniques like focused ion beam milling and electron beam lithography. These structures are nanosized, up to $1000\,m$, and more particularly up to a few hundred nanometers.[8]

**Sensing and Enhancement:** The nanoscale size and localized fields of these features are the driving force behind enhanced detection.

- **Hot Spots:** Nanostructures support strong electromagnetic field enhancement or optical "hot spots". These are crucial for Surface-Enhanced Raman Scattering (SERS) down to the single-molecule level.[42], [54], [62], [66], [67]

- **Antennas:** Plasmonic nanoantennas (e.g., bowties, nanorods, dipole antennas, nanocrescent antennas) enhance the detection and emission of plasmons. They act as antennas, focusing energy to a tiny focal spot.[14], [30], [31], [45], [48], [50], [67], [68], [69], [70]

- **Quantum Coupling:** Nanostructures are necessary to observe quantum effects, such as metal-enhanced fluorescence of colloidal nanocrystals and Rabi splitting from strong plasmon-molecule coupling within plasmonic cavities or aluminum nanodisk arrays.[11], [14], [15]

- **Lithographic Manufacturing:** The complexity of nanostructures necessitates advanced fabrication techniques.

- **High-Resolution Fabrication:** Creating highly confined plasmonic structures, such as nanogaps and intricate patterns, typically relies on lithography techniques like electron-beam ($e-beam$) lithography, focused ion-beam lithography, photolithography, nanoimprinting (including roll-to-roll), or nanosphere lithography.[2], [7], [8], [53], [59], [71], [72], [73]

- **Emerging Techniques:** Plasma-assisted nanofabrication is noted as an emerging method for creating nanowires, nanotubes, nanoparticles, and nanotextured coatings. Alternative methods, such as light-directed reversible assembly using plasmon-enhanced thermophoresis, enable dynamic lithography to form arbitrary patterns of dispersed nanoparticles at the solid-liquid interfaces.[27], [42]

## 1.3   Plasmonic Waveguides: Structure, Function, and Integration

Plasmonic waveguides are pivotal in manipulating surface plasmon polaritons (SPPs) and light at the nanoscale, enabling subwavelength confinement and guidance that surpasses the classical diffraction limit. These structures are integral to applications in photonic circuits, sensing, and heat-assisted magnetic recording (HAMR). This document details the structure, functionality, and integration of plasmonic waveguides.

### 1.3.1   Role and Characteristics of Plasmonic Waveguides

**Plasmonics** (the field dealing with surface plasmon waves) focuses on phenomena like making waveguides for these waves, modulating them, and studying related sources, receivers, and couplers.[18]

- **Subwavelength Confinement and Guiding**: Plasmonics offers a way to guide light below the diffraction limit while still maintaining a high optical bandwidth, providing a different solution for nanoscale light wave processing. Waveguides can be fabricated to utilize resonant plasmonic energy with relatively long coherence to transmit optical energy or information over a distance.[8], [45]

- **SPP Propagation:** Plasmonic waveguides are systems realized to form the building blocks of a chip-based plasmonic system. The Surface Plasmon Polariton (SPP) waveguide is designed to guide SPP signals. SPP waveguides may include a metal line or double metal line, typically having a thickness of ones to tens of nm and a width ranging from $10\,nm\,to\,10\,\mu m$.[8], [18]

- **Trade-offs:** The coupling of propagating light to electron oscillations (plasmons) enables extreme modifications to the light but comes at the cost of enhanced attenuation (high intrinsic loss due to resistive heating in metals). This intrinsic metal loss limits the quality factor of plasmonic devices.[14], [74], [75]

## 1.3.2 Types of Waveguide Structures

The sources describe several configurations for plasmonic waveguides:

### A. Hybrid Plasmonic Waveguides (HPWs)

HPWs combine the advantages of surface plasmon confinement with the long-range propagation of dielectric modes.[14]

- **Function and Composition:** HPWs involve new modes arising from the hybridization between surface plasmons and dielectric waveguides. They are known for providing both subwavelength confinement of light and long-range propagation.[14], [28], [76], [77], [78], [79]

- **Applications:** HPWs are relevant for devices that require long-range guiding of surface plasmons with strong subwavelength confinement. They are crucial for studying extreme interactions between light and matter in active photonic circuits, bio-sensing, and quantum information technology.[14], [80]

- **Specific Structures:** HPWs can be formed, for instance, by separating a high-gain semiconductor nanostructure (e.g., CdS nanowire) from a metal surface (e.g., silver) by a nanoscale low-index gap (e.g., $5 - nm\ thick\ insulator$). They can also be hybrid metal-insulator rib waveguides with subdiffraction limit mode confinement. Examples include the hybrid plasmon-waveguide modes (HPWMs) used in optically rewritable photonic waveguides.[12], [14], [80]

### B. Slot Waveguides

These structures confine light in a narrow dielectric gap between two metal surfaces, utilizing the plasmonic gap mode.[68]

- **Structure:** Common types include Metal-Insulator-Metal (MIM) SPPs and Insulator-Metal-Insulator (IMI) SPPs. MIM slot waveguide structures (e.g., $Ag/SiO_2/Ag$) can be used to guide light between two metal layers.[12], [34]

- **Confinement and Intensity:** The plasmonic nature of the slot waveguide provides extreme sub-diffraction confinement of the optical field, allowing electrode spacing on the order of 100 nm (or even 10 nm to 50 nm in certain photodetectors). This strong confinement leads to very strong optical intensities, essential for strong nonlinear effects and terahertz (THz) devices.[81], [82]

- **Applications:** Slot waveguides are integral to plasmonic modulators, such as those using Transparent Conducting Oxides (TCO). They are also used in THz generation and detection devices, often integrated with antenna arms.[68], [82]

## C. Nanowire and Other Simple Waveguides

Plasmonic waveguides can also take the form of simple wire structures for chip-based systems.[8]

- **Nanowire Waveguides:** Materials such as Ag nanowire(s), Au nanowire(s), or Cu nanowire(s) are nonlimiting examples of plasmonic waveguides. Nanowire systems can act as waveguides for 1D-surface plasmons at optical frequencies.[9], [38], [77]

- **Channel and Groove Waveguides:** These are noted as configurations of interest, along with stripe, wedge, and slot waveguides.[8], [12], [83]

## 1.3.3 Integration and Functional Components

Waveguide structures are essential for coupling light into or out of plasmonic components and routing signals within nanoscale devices.

## A. Optical Coupling and Mode Conversion

Waveguides are designed to convert standard optical signals into highly confined plasmonic modes or vice versa:

- **Coupling Mechanisms:** Optical coupling (or evanescent-wave coupling) is fundamentally identical to near-field interaction, achieved by placing two or more electromagnetic elements (like optical waveguides) close together so the evanescent field of one reaches the other.[84]

- **Tapered Waveguides and Coupling Units:** To connect conventional optical waveguides (micrometer size) to nanoscale SPP waveguides, special coupling units are required. These include dielectric taper waveguides or input double metal tapers. These units convert the optical mode into the SPP mode and focus the signal to micron or nano sizes.[18]

- **Focusing Structures:** Nano SPP waveguide type input focusing devices, such as tapered channel metal waveguides, tapered needle metal waveguides, and Bragg grating metal waveguides, function to gradually reduce the SPP mode to a nano size and transfer it to the SPP waveguide.[18]

## B. Application in Integrated Circuits (ICs)

Plasmonic waveguides serve as key interconnects in integrated devices:

- **Chip-Scale Integration:** Plasmonic waveguides enable the realization of highly integrated photonic circuits having a micro structure, low power consumption, and low price.[18]

- **Electronic-Photonic Integrated Circuits (EPICs):** In EPICs, the waveguide acts as a photonic or plasmonic bridge to transmit signals between emitters and photodetectors. Plasmonic waveguides provide an optical link between transistors, replacing traditional copper links.[9]

- **Modulators and Sensing:** Waveguides are core elements in various plasmonic devices, including plasmonic phase modulators, which launch SPP waves onto a plasmonic layer (e.g., graphene sheet). Waveguides in nano plasmonic circuits can also incorporate signal sensing/processing units, sometimes exposed to a cladding layer to operate in response to refractive index changes. Examples include waveguides configured as Y splitters or Mach-Zehnder interferometers.[18], [85]

## C. Application in Heat-Assisted Magnetic Recording (HAMR)

In HAMR systems, waveguides are used to supply light to the Near-Field Transducer (NFT).

- **Light Delivery:** A channel waveguide core delivers light to the NFT, which then generates surface plasmon-enhanced, near-field electromagnetic energy.[86]

- **Structural Elements:** HAMR devices use waveguides alongside components like waveguide blockers (WGBs) and optical side shields (OSS) for controlling light delivery. The waveguide core often narrows as it approaches the media-facing surface.[26], [87]

- **Materials for Stability:** For mechanical stability, especially near the output surface where high temperatures occur, designs often employ weakly plasmonic materials (like Rh, Ir, Pt, Pd) in elements like the waveguide blocker, optical field enhancer, and optical side shields, where they support weaker plasmon modes but provide robustness, while highly efficient (but soft) plasmonic materials (like Au) are recessed away from the critical surface.[26], [87], [88]

## 1.3.4 Materials Supporting Plasmonic Waveguides

Plasmonic materials are selected for their ability to support SPPs.[12]

- **Noble Metals:** Ag, Au, and Cu are commonly cited materials for plasmonic waveguides. Gold (Au) or Silver (Ag) surfaces separated from a high refractive index medium (usually silicon) by a small physical gap often form the confinement structure in plasmonic waveguides.[9], [12], [46], [74], [88], [89]

- **Graphene and TCOs:** Graphene sheets are used as plasmonic layers in phase modulators. Transparent conducting oxides (TCOs), such as indium tin oxide (ITO), are used in plasmonic devices, sometimes near their epsilon-near-zero (ENZ) region, which can be dynamically tuned electrically or non-linearly optically.[8], [68], [85]

- **Non-Metallic Materials:** The term "plasmonic layer" is not strictly limited to metals and can be formed from non-metallic materials, especially those selected for high reflectivity at the wavelength of interest. Vanadium oxide ($VO_2$) and chalcogenide films (e.g., $Ge_2Sb_2Te_5$) are examples of solid-state phase change materials used in plasmonic waveguides/modulators that transition between insulating and metallic phases.[12]

# 2 Fundamental Plasmonic Effects

## 2.1 Resonance Types

Plasmonic effects arise from the collective oscillation of free electrons in conductive materials, typically metals, forming quantized excitations known as plasmons. These oscillations couple with incident electromagnetic (EM) fields, such as photons, to form quasiparticles like surface plasmon polaritons (SPPs). This document outlines the primary plasmonic resonance types and their associated fundamental effects.

### 2.1.1 Surface Plasmon Resonance (SPR) / Propagating Surface Plasmon Polaritons (SPPs)

**SPR** refers to the physical phenomenon where incident light stimulates collective electron oscillations at the metal surface for planar surfaces.[1], [90], [91]

- **Nature and Propagation:** SPR involves the excitation of surface plasmon polaritons (SPPs), which are electromagnetic surface waves that propagate along the interface between a metal and a dielectric (or vacuum/air). SPPs are longitudinal EM waves arising from the interaction of light with free electrons. They propagate parallel to the interface for distances on the order of microns to tens or even hundreds of microns. Their field strength decreases exponentially (decays evanescently) normal to the surface.[2], [90], [91], [92], [93], [94]

- **Excitation Conditions:** Excitation requires momentum matching between the photon and the plasmon. Since the SPP wave vector is larger than that of the incident radiation, auxiliary coupling techniques are needed.[90], [92], [95], [96]

  ◦ **Kretschmann Configuration:** This common method involves coupling excitation light through a high-refractive-index prism adjacent to a thin, SPR-active metal layer. The resonant conditions are affected by the refractive index on both sides of the metal surface, which is used in biosensors to detect biological interactions on top of the metal surface. SPR is responsible for a dip in reflectance at a specific wavelength due to the absorption of optical energy in the metal.[90], [91], [92], [95], [97], [98]

  ◦ **Grating Coupling (GCSPR):** This method replaces the prism with a diffraction grating (e.g., grooves, slits, holes, or arrays of scatterers) patterned on a flat metal surface. The grating couples higher-order diffracted light into the surface plasmon to achieve resonance.[92], [95], [96], [97], [99]

  ◦ **Polarization:** Excitation of SPPs typically requires p-polarized light (TM polarization) because waves propagating along the interface must have an electric field component perpendicular to the surface.[90], [95]

## 2.1.2 Localized Surface Plasmon Resonance (LSPR)

**LSPR** refers to the surface plasmon resonance of nanometer-sized structures, such as metallic nanoparticles.[1], [53], [62], [91], [92], [100]

- **Confinement:** LSPR occurs when the oscillating electrons are spatially confined within metallic nanostructures with lateral dimensions generally less than half the wavelength of the exciting EM wave.[58], [92], [101]

- **Resonant Response:** The confinement results in a characteristic plasmon frequency at which there is a resonant response to the incident field. The plasmonic nanostructure achieves maximum absorption at this resonance frequency or wavelength, often resulting in strong UV-V is absorption bands.[58], [92], [101], [102]

- **Field Enhancement and Hot Spots:** LSPR induces greatly enhanced electric (E) fields in confined nanoscale locations, known as "optical hot spots" or "hot spots," typically on the tips of nanorods or in the junctions/gaps of nanodimers/clusters. These enhancements can be enormous (up $to\ 10^{15}$ −fold for SERS).[37], [38], [42], [52], [53], [58], [60], [62], [67], [101], [103], [104], [105]

- **Tunability:** The resonance wavelength of LSPR nanostructures is tunable based on design parameters such as the size, shape, and architecture of the nanostructure, as well as the local dielectric environment/refractive index of the surrounding medium. For example, gold and silver nanorods can exhibit two peaks: one due to lateral modes and one due to longitudinal modes, tunable from the visible through the near infrared based on the nanorod aspect ratio.[57], [102], [104], [105], [106], [107]

- **Coupling Conditions:** Unlike SPR, LSPR coupling conditions are less stringent and do not depend on the incident angle, which simplifies coupling into an LSPR particle.[97], [108]

### 2.1.3 Long-Range Surface Plasmon Resonance (LRSPR)

**LRSPR** is a type of plasmon specific to thin metal films or strips.[92]

- **Propagation:** It is characterized by low attenuation and can travel along the surface for distances up to a few millimeters in the visible spectrum, or even a few centimeters in the infrared.[92]

- **Applications:** This type of plasmon might be useful in active photonic components and highly sensitive sensors, particularly for large biological entities like cells.[92]

### 2.1.4 Specialized and Engineered Resonance Types

The sources also describe several highly specialized or engineered resonance mechanisms:

- **Toroidal Resonances:** These modes are distinct from classical electric dipolar and multipolar moments and are associated with a unique family of dynamic toroidal multipolar modes arising from oscillating radial components of the current density. They have been observed in 3D structures but are also studied in planar 2D structures. Excitation in planar structures can lead to the formation of a closed-loop head-to-tail magnetic moment configuration around a central part of a unit cell. Toroidal resonances can provide a high-Q factor and are polarization-dependent, allowing for applications in biosensing and THz switching.[49]

- **Spoof Plasmonics and Hybrid-Spoof Plasmonics:** Spoof plasmonics involve defining surface geometries on metallic conductors to mimic the dispersion and confinement of SPPs. This can be achieved with any conductor (or highly doped semiconductors/polar dielectrics) at frequencies below the plasma frequency, with the operating frequency defined primarily by the geometry. Hybrid-spoof plasmonics combine traditional plasmonic materials (like Ag or Au) with lithographically defined geometric patterns to induce spoof-plasmon excitations, offering further control of SPPs and/or multiple resonant structures. This mechanism allows for realizing spoof-plasmon resonances in the visible spectrum, which is difficult with standard lithography due to the required small spacing between grains created by the deposition process (e.g., Atomic Layer Deposition).[109]

- **Exciton-Plasmon Coupling/Hybridized Modes:** Plasmonic coupling can be either weak or strong. When the interaction is strong (the strong coupling regime), it leads to the formation of new quasiparticles from admixtures of strongly coupled states, which can exhibit unusual properties. Examples include strong coupling between a surface plasmon mode and organic excitons, which can be leveraged in Plasmonics-Enhanced Upconversion (PE-UC) or Exciton-Plasmon Enhanced Phototherapy (EPEP) probes. This interaction can result in phenomena like Rabi splitting.[14], [38], [58], [98]

- **Epsilon-Near-Zero (ENZ) Modes:** These are bulk plasmon modes that occur at the ENZ frequency in ultrathin plasmonic materials. In specific configurations, an external plane wave can couple to the ENZ mode (sometimes called the Berreman mode or ENZ mode depending on the coupling location) and be completely absorbed in a deeply sub-wavelength film. The field enhancement properties of the ENZ mode can lead to substantial increases in nonlinear optical processes like harmonic generation.[98]

- **Leaky Modes and Fano Resonances:** Leaky modes are resonant leaky modes in periodically modulated films, whose resonance locations can be precisely controlled. In some systems, sharp, asymmetric Fano resonances arise from interference between a continuous background (broad resonance) and a discrete state (narrow resonance). Relatedly, a phenomenon analogous to Electromagnetically Induced Transparency (EIT) develops when two weakly localized modes (like a Plasmonic Detector (PD) mode and a Tamm state) are coherently coupled and resonate at the same frequency with asymmetric damping rates.[49], [64], [110]

- **Chiral Plasmonic Modes:** These are elicited by 3-D metal dielectric structures with chiral symmetry. The resulting plasmons may exhibit propagational circular polarization, whose frequency can be tuned to couple with electronic transitions or fundamental vibrations of a molecular species having spatial chirality.[94]

**Summary of Fundamental Plasmonic Effects**

Regardless of the specific resonance type, the fundamental plasmonic effects utilized across these applications include:

1. **Concentration and Confinement of EM Waves:** Plasmons concentrate EM waves in regions considerably smaller than their wavelength, allowing for sub-wavelength confinement of light and control of optical fields.[1], [2], [3], [53], [92]

2. **Field Enhancement:** This is the resulting increase in local electric fields, particularly at "hot spots". This enhancement is key to techniques like Surface-Enhanced Raman Scattering (SERS) and metal-enhanced bioluminescence.[2], [37], [38], [52], [53], [60], [101], [103], [111]

3. **Refractive Index Sensitivity:** The resonance conditions (frequency/wavelength/angle) are extremely sensitive to changes in the local dielectric environment or refractive index near the metal-dielectric interface. This is the basis for SPR and LSPR biosensors.[1], [90], [92], [105], [106], [108], [112], [113]

4. **Spectral Tunability:** The resonance frequency or wavelength can be tailored by controlling the material composition, size, shape, and architecture of the plasmonic structures.[102], [105], [106]

5. **Thermal Effects (Photothermal):** Resonant light absorption accompanying plasmon excitation generates local heat (thermal hot spots) within the nanostructures, which is exploited in thermoplasmonics, optothermal tweezers (OTTs), and electrothermoplasmonic manipulation.[42], [62], [69], [114]

6. **Energy Transfer:** Plasmons can facilitate the transfer of energy, such as non-radiative coupling of bioluminescent emission to surface plasmons, or coupling of plasmonic optical power to electrical power (plasmoelectric effect).[102], [111]

## 2.2  Plasmonic Wave Phenomena and Fundamental Effects

Plasmonic wave phenomena arise from the interaction of light (photons) with collective electron oscillations (plasmons) in conductive materials, forming hybrid electromagnetic waves that enable subwavelength confinement and manipulation of light. This document details the key wave phenomena and their associated fundamental plasmonic effects.

### 2.2.1  Surface Plasmon Polaritons (SPPs) and Plasmon Propagation

Surface plasmons (SPs), also known as Surface Plasmon Polaritons (SPPs), are the fundamental electromagnetic (EM) waves inherent to plasmonics.[1], [17], [46], [90], [91], [92], [94]

- **Nature of SPPs:** SPPs are collective oscillations of free electrons at a metal/dielectric (or metal/vacuum) interface, emerging as longitudinal EM waves. Plasmons are defined as the quanta of these plasma oscillations, considered quasiparticles that arise from the quantization of plasma oscillations, similar to how phonons are quantizations of mechanical vibrations.[1], [46], [92], [94]

- **Propagation and Confinement:** SPPs are electromagnetic surface waves that propagate parallel to the interface. Their field strength decreases exponentially (decays evanescently) normal to the surface. This allows SP waves to be tightly bound to the interface, penetrating around 10 nm into the metal and typically more than 100 nm into the dielectric. Critically, SPPs concentrate EM waves in a region considerably smaller than their wavelength. They can propagate for distances on the order of microns to tens and even hundreds of microns.[1], [90], [92], [94]

- **Long-Range Propagation:** A specialized version, Long Range SPR (LRSPR), exists in thin metal films or strips and is characterized by low attenuation, capable of traveling along the surface for distances up to a few millimeters in the visible or a few centimeters in the infrared.[92]

- **Wave Guiding:** Plasmons are utilized to transmit optical energy or information over a distance via waveguides. A plasmonic waveguide is a cavity configured to achieve strong plasmon confinement. Hybrid plasmonic waveguides enable subwavelength confinement and long-range propagation.[28], [45], [46], [77], [85]

### 2.2.2  Excitation and Wavevector Matching

Excitation of SPPs requires matching the momentum (wave vector) of the incident photon with that of the plasmon.[90], [92], [94]

- **Momentum Mismatch:** The SPP wavenumber ($k_{SPP}$) is generally larger than that of the incident radiation ($\omega/c$), meaning the incident light alone is insufficient to excite the mode.[6], [90], [95]

- **Coupling Techniques:** Auxiliary coupling techniques are necessary to enhance the wave vector and achieve phase-matching.[6], [90], [92], [95]

◦ **Kretschmann Configuration:** Uses a high-refractive-index prism adjacent to the metal film. An evanescent electric field extends through the metal film, exciting plasmons at the outer interface. This configuration is also exploited in Epsilon-Near-Zero (ENZ) modes excitation by coupling light through a glass block under total reflection conditions.[90], [91], [92], [95], [98], [115]

◦ **Grating/Aperture Coupling:** Uses a diffraction grating or arrays of nanoapertures to enhance the wave vector by diffraction. In patterned metallic films, light can diffract and couple into surface plasmon modes.[1], [6], [40], [92], [95], [99]

◦ **Polarization:** SPP excitation typically requires p-polarized light (Transverse Magnetic, TM polarization) because waves propagating along the interface must have an electric field component perpendicular to the surface.[90], [95], [116]

## 2.2.3  Wave Confinement and Localized Effects

In nanostructures, wave phenomena become highly localized, leading to immense field enhancement.

- **Localized Surface Plasmon Resonance (LSPR):** Occurs in metallic nanostructures (nanoparticles) with lateral dimensions typically less than half the exciting EM wave wavelength. The oscillating electrons are spatially confined, leading to a resonant response.[60], [92], [105]

- **Enhanced Electric Fields and Hot Spots:** LSPR induces greatly enhanced electric (E) fields in confined nanoscale locations, known as "optical hot spots". This phenomenon is the basis for techniques like Surface-Enhanced Raman Scattering (SERS), where the field enhancement factor is often proportional to the fourth power of the local electric field.[53], [60], [62], [101]

- **Subwavelength Focusing:** Plasmonics allows for subwavelength confinement of light. For example, plasmonic micro-zone plates (PMZPs) can focus beams as small as $\sim$ 220 $nm$.[2], [75], [117]

## 2.2.4  Coupled and Hybrid Wave Modes

Plasmonic effects often involve the coupling of different types of waves, forming hybrid modes or complex wave interactions.

- **Hybrid Photonic-Plasmonic Modes:** These modes are crucial in several device concepts.

◦ **Waveguide Modes:** In plasmonic quantum well lasers, a hybrid photonic-plasmonic mode is likely the most favorable for relatively thin gain regions ($\sim 20 - -100 \ nm$). The plasmonic waveguide "anchors" the optical mode to the small region of the metal interface, while emission occurs in a radiative photonic mode.[46]

◦ **Surface Lattice Resonances (SLRs):** In ordered arrays of optical antennas (nanoparticles), localized SPPs couple via diffraction, leading to collective, lattice-induced, hybrid photonic-plasmonic resonances known as SLRs. These modes are delocalized, extending over several unit cells.[118], [119]

- **Exciton-Plasmon Coupling:** Strong interaction between plasmon modes and electronic states (excitons) leads to the formation of new quasiparticles (mixed plasmon-exciton states) from the admixture of strongly-coupled states. This strong coupling regime may result in phenomena like Rabi splitting.[14], [37], [38], [41], [52], [58]

- **Epsilon-Near-Zero (ENZ) Modes:** These bulk plasmon modes occur at the frequency where the real part of the material's dielectric constant crosses zero. In ultrathin films, external light can couple to this volume plasmon (sometimes called the Berreman mode or ENZ mode) and be completely absorbed, resulting in large electric field enhancement.[98]

## 2.2.5 Complex Wave Interactions and Interference

The field also exploits complex wave behaviors like interference and topological effects.

- **Plasma Waves (Non-Optical Frequencies):** The sources discuss plasma waves in 2D electron fluids (e.g., in field effect transistors) used for detecting and emitting radiation in the THz to far infrared range. These plasma waves are oscillations of electron density varying in time and space.[120]

◦ **Plasmonic Crystals (PCs):** Introducing spatial periodicity to a 2DEG (two-dimensional electron gas) creates a plasmonic crystal that exhibits band structure. The wave coherence across the PC lifts degeneracy, resulting in the formation of state bands.[64]

◦ **Tamm States and Defect Modes:** Plasmonic crystals support specialized wave states such as Tamm states (weakly localized crystal surface states typically found in a band gap) and Plasmonic Defect (PD) modes.[64]

◦ **Induced Transparency (EIT Analog):** An effect analogous to Electromagnetically Induced Transparency (EIT) can develop when two weakly localized modes (like PD and Tamm states) are coherently coupled and resonate at the same frequency with asymmetric damping rates.[64]

- **Fano Resonances:** Plasmonic systems can exhibit Fano-type resonances, resulting from the interference between a continuum (e.g., allowed bands of a PC) and a discrete mode (e.g., a cavity mode). Sharp, high-Q factor Fano resonances are of significant interest.[49], [64]

- **Electromagnetic Vortices and Polarization:** Engineered structures can generate specific topological wave phenomena.
  ◦ **Optical Vortices:** These are electromagnetic fields possessing a dot-like phase singularity, around which the phase is wrapped, creating helical phase fronts. An optical vortex carries a non-null orbital angular momentum (OAM). Plasmonic plates can be designed to generate these optical vortices by coupling surface plasmons and air, where the coupling enables the light propagation through the plate.[121]

  ◦ **Chiral Plasmonic Modes:** 3-D metal dielectric structures with chiral symmetry elicit SPPs that may exhibit propagational circular polarization. The vortices associated with the localized electric fields of these SPPs derive from the handedness of the external field and the chiral structure.[94]

In summary, fundamental plasmonic effects are deeply rooted in the wave nature of light and matter, leveraging the ability of plasmons to propagate as highly confined, short-wavelength surface waves (SPPs) and to participate in complex resonance, coupling, and interference phenomena in nanostructured or periodic materials.

## 2.3 Manipulation and Tuning of Plasmonic Effects

The manipulation and tuning of plasmonic effects are central to harnessing surface plasmons for dynamic applications in sensing, signal processing, lithography, and energy management. These methods enable precise control over plasmon resonance properties, energy flow, and physical object manipulation. This document categorizes the manipulation and tuning techniques into three primary areas: resonance frequency and optical property tuning, energy and power flow manipulation, and dynamic physical object manipulation.

### 2.3.1 Tuning the Resonance Frequency and Optical Properties

The most common form of tuning involves controlling the plasmon resonance frequency ($\omega_p$) and the associated optical response (phase, amplitude, absorption).

## A. Geometric and Material Tuning (Passive/Static)

Plasmonic nanostructures offer high spectral tailorability, allowing their resonant behavior to be set during fabrication.

1. **Nanostructure Geometry and Size:** The plasmonic resonance of nanostructures can be tuned from the ultraviolet through the visible and infrared spectrum based on design parameters such as the size, shape, and architecture of the nanostructures.[102]

   ◦ **Localized Surface Plasmon Resonance (LSPR):** The excitation conditions for LSPR are determined by the size and shape of the particle or pattern element in relation to the excitation wavelength.[91]

   ◦ **Aspect Ratio and Shape:** For prolate structures (like nanorods or ellipses), tuning depends on the ratio of the long axis to the short axis (aspect ratio). Gold and silver nanorods exhibit tunable longitudinal modes in the visible through near-infrared spectrum depending on the aspect ratio. For spheroidal shells, the shell thickness and the shell aspect ratio provide two degrees of freedom for tuning the plasmon resonances. Rectangular nanostructures can support multiple resonant modes with different frequencies depending on their differing length and/or width.[37], [38], [52], [91], [103], [106], [122]

   ◦ **Refinement of Features:** The spot size of a plasmonic micro-zone plate (PMZP) can be reduced by optimization of its geometrical parameters. The spectral response of a plasmonic structure can be tuned by selecting its form factor, periodicity, and metallization. Varying the dimensions and/or orientation of plasmonic nanoantennas can vary the relative intensity of the optical response in a given spectral band or polarization.[117], [123], [124]

2. **Material Composition:** The optical properties (absorption/scattering) can be tailored by varying the composition of the metallic nanostructures. Tuning the composition or size of the plasmonic material in multicomponent photocatalysts allows for photocatalysis at specific wavelengths.[5], [113], [125]

3. **Grating Periodicity and Wavelength:** In grating-coupled systems, the spectral response (resonance peak location and width) can be varied by changing the periodicity (pitch) of the nanostructures, which affects radiative coupling and interference of scattered light. The resonance frequency of a plasmonic device can be controlled by changing a layer thickness, which shifts the dispersion curves of the propagating modes.[1], [12], [99], [108]

**B. Active Tuning Mechanisms (Dynamic)**

Dynamic tuning utilizes external stimuli (voltage, strain, light, heat) to modulate the plasmonic response after fabrication.

1. **Electrical and Electro-optical Modulation:** This is achieved primarily by altering the dielectric environment or charge carrier concentration.

   ◦ **Refractive Index Modulation via Voltage (Liquid Crystals/Electro-optics):** Applying a bias voltage to plasmonic waveguides containing an electrically-adjustable dielectric (like liquid crystal or electro-optic polymers) dynamically modulates the refractive index and associated permittivity. This small change in refractive index of a high-Q plasmonic resonant waveguide leads to a substantial shift in the resonant wavelength(s), which can achieve nearly $2\pi$ phase modulation. Such devices can be used for beamforming by mapping voltage differentials to unique reflection phases across an array.[89]

   ◦ **Refractive Index Modulation via Charge Injection (Solid State):** Applying a voltage across a thin solid electrolyte layer can drive protons into the layer, changing its refractive index, which can be used to implement switchable plasmonics.[126]

   ◦ **Phase Change Materials:** Plasmonic modulators can use solid-state phase change materials ($e.g., vanadium oxide, Ge_2Sb_2Te_5$) that transition between an insulator phase (transparent) and a metallic phase (opaque) when triggered by an electrical field, heat, or ultrafast optical pumping. This alters the available modes of propagation, allowing for index (phase) modulation, amplitude modulation, or waveguide mode cutoff.[12]

   ◦ **Controlling Carrier Density in Semiconductors/Graphene:** The plasmon frequency ($\omega_p$) in a semiconductor or graphene is fundamentally dependent on the concentration of free charge carriers (N). This concentration can be influenced by doping during fabrication or by injection of free charge carriers under operation via a tuning control voltage (gating) . Examples include:[57], [85], [93], [127]

   ◦ **p-i-n structures:** Controlling the free charge carriers in an intrinsic semiconductor layer through a control voltage applied in forward bias can increase the density of carriers, varying the plasmon resonance frequency.[93]

   ◦ **Graphene/TCO Tuning:** Applying a bias voltage to a graphene plasmonic layer controls its Fermi energy ($E_F$), which, in turn, modulates the propagation speed and phase of the SPP wave. Similarly, gate-tunable Epsilon-Near-Zero (ENZ) materials like Transparent Conducting Oxides (TCOs) can electrically and nonlinearly tune plasmonic nanocircuits for advanced light manipulation, controlling optical confinement, phase, and amplitude.[8], [85]

▪ **Controlling Effective Mass in Semiconductors:** The plasmon frequency is also dependent on the effective mass (m) of the free charge carriers. The effective mass can be influenced by subjecting the crystal lattice of the plasmon resonance layer to mechanical stress (compressive or tensile stress). This stress can be applied piezoelectrically via a stressor layer, which changes the lattice strain and subsequently the effective mass, tuning the plasmon frequency.[93], [127]

2. **Mechanical/Thermal Tuning:**
   ◦ **Lattice Strain:** Varying lattice stress in a semiconductor plasmon resonator is a way to alter the resonance frequency by changing the effective mass of the free charge carriers.[93]

   ◦ **Geometry Alteration:** The resonance frequency can be subsequently tuned by altering the geometry of the resonator.[93]

   ◦ **MEMS/NEMS:** Mechanical mobility in nanostructured systems can realize performance enhancements. Tunable resonant leaky-mode $N/MEMS$ elements can achieve spectral tuning by mechanically altering the coupling of guided-mode resonances. In Plasmonic Microelectromechanical (PMEM) devices, applying body-bias to piezoelectric actuators can modulate the size of the initial air gap, enabling the trimming and reconfiguration of the IR sensitivity and threshold of the device.[110], [128]

## 2.3.2  Manipulation of Energy and Power Flow

Plasmonic effects are used to actively direct, diffuse, or control the movement of energy (both electromagnetic radiation and thermal energy).

1. **Power Diverting/Thermal Management (Fractal Plasmonic Surfaces - FPS):** Fractal plasmonic surfaces (FPS's) are used to transfer radiation, such as via evanescent wave transfer, to remove radiative power and/or heat from one location to another, or divert it to another location. An FPS can diffuse power delivered to a localized "hotspot" and spread it out or divert it to other desired locations. FPS on curved surfaces may act as a cloaking device to divert power in a frequency of interest from one side to the other.[129]

2. **Optical Steering and Filtering:**
   ◦ **Beam Steering/Holography:** Dynamic metasurfaces, composed of arrays of adjustable plasmonic resonant waveguides, enable optical beamforming (steering) in both transmit and receive applications by applying voltage patterns to control the reflection phase of the elements.[89]

◦ **Filtering and Switching:** Plasmonic waveguides and nanocircuits are engineered to function as filters and switches. In optical fibers, plasmonic devices can implement a fast pedestal suppression function (passive optical valve) and other complex transmission shapes like a saturable absorber function by controlling the nonlinear optical coupling scheme between the fiber core and the plasmonic device.[12], [74], [127]

3. **Active Plasmoelectric Control:** The plasmoelectric effect permits active control of the Fermi level of a metal using incident radiation. Furthermore, this effect can modulate the absorption cross section at specific frequencies for applications relating to optical switching.[102]

4. **Polarization Control:** Plasmonic toroidal resonators exhibit polarization dependency, which can be exploited for fast and efficient on/off routing and filtering purposes by rotating the incident magnetic field angle. Chiral plasmonic structures produce SPPs with propagational **circular** polarization whose frequency is tuned by design to couple selectively with molecules having spatial chirality.[49], [94]

### 2.3.3 Dynamic Manipulation of Physical Objects (Trapping and Transport)

Plasmonic effects are leveraged to remotely manipulate particles and biological cells in fluid media, primarily using photothermal effects.

1. **Optothermal Tweezers (OTTs):** Plasmon-enhanced photothermal effects are used to create systems known as optothermal tweezers (OTTs).[42], [114]

◦ **Mechanism:** Absorbing light at the plasmon resonance generates localized heat (thermal hot spots) within the plasmonic substrate. This local heating creates temperature gradients ($\nabla T$) in the surrounding fluid, which induce electrothermal flows, convective drag, and thermophoretic forces that capture and confine particles.[56], [62], [114], [130]

◦ **Control and Versatility:** The confinement region's diameter can be controlled by the power density of the electromagnetic radiation. The optothermal tweezers allow versatile manipulation of biological cells and nanoparticles at moderate temperature rise and gradient. The systems can achieve trapping speeds ranging from $200\ nm/s\ to\ 50\ \mu m/s$.[56], [114]

◦ **Arbitrary Patterning:** A Digital Micromirror Device (DMD) or Spatial Light Modulator (SLM) can be used to control the optical images/patterns, enabling the simultaneous trapping and arbitrary manipulation of colloidal particles or biological cells.

These patterns can be "erased" and "rewritten" repeatedly (dynamic lithography).[42], [114]

2. **Electrothermoplasmonic (ETP) Flow:**
   This hybrid approach combines plasmon-induced heat with an applied AC electric field.

   ◦ **Rapid Transport:** ETP flow transports suspended particles rapidly and precisely toward plasmonic trapping sites (hotspots). This process overcomes the slow, diffusion-limited transport associated with purely optical gradient forces, especially for individual nano-objects.[56], [69]

   ◦ **High-Resolution Trapping:** By illuminating a single nanoantenna and applying an AC field, high-resolution stable trapping of individual particles at a given hotspot is achieved, overcoming the particle agglomeration issues faced by array illumination techniques.[69]

3. **Selective Functionalization:**
   The tuning process can be used to functionalize specific regions of the plasmonic substrate for selective capture.

   ◦ **Selective Coating:** Plasmonic components can be selectively coated with polymers to change their local dielectric environment and resonance properties. Selective tuning of resonance can be achieved by capturing polymers of different optical properties and heating them to coat selected nanoantennas, which can be used to achieve color printing.[130]

# 3  Applications

## 3.1  Plasmonic Detection Principles for Biosensing and Diagnostics

Plasmonic detection principles leverage the high sensitivity of surface plasmons to enable label-free, real-time, and high-throughput detection of chemical and biological analytes, often achieving single-molecule resolution. These principles underpin advanced biosensing and diagnostic applications. This document categorizes the detection mechanisms into three primary areas: resonance shift and refractive index sensing, enhancement of emission and spectroscopy, and multi-modal and advanced detection schemes.

### 3.1.1 Resonance Shift and Refractive Index Sensing

This is the fundamental, label-free mechanism of detection, where the binding of an analyte perturbs the electromagnetic environment near the plasmonic structure, shifting its characteristic resonance.

**A. Surface Plasmon Resonance (SPR)**

In conventional SPR and related methods, the presence of an analyte is detected by monitoring changes in the refractive index of the adjacent medium.[90], [96]

1. **Sensing Mechanism:** The excitation condition for surface plasmon polaritons (SPPs) is extremely sensitive to changes in the dielectric parameters (refractive index) of the medium immediately surrounding the metal surface. When analyte molecules bind to linkers (capture agents) functionalized on the metal film, the local refractive index changes.[90], [96], [131]

2. **Detection Modalities:**

   ◦ **Angular Modulation (AM):** Measures the reflectivity as a function of the angle of incidence using a single wavelength, typically a collimated laser beam. The characteristic SPR dip is observed in the reflectivity spectrum as the angle of incidence is scanned. The angle at which the minimum intensity is present is displaced when analyte molecules couple to the metallic film.[92], [132]

   ◦ **Wavelength Modulation (WM):** Measures the reflected power for a given angle of incidence by varying the wavelength of the incident light. A resonant wavelength shift is monitored.[90], [92]

   ◦ **Imaging Schemes (SPR Imaging):** Used to increase throughput, allowing several device cells ($\sim 10^2$) to be used in parallel to image the binding interaction and monitor intensity variations caused by a refractive index change in each cell.[96]

3. **Enhanced SPR Architectures:**
   Advanced designs significantly boost sensitivity and depth of detection:

   ◦ **Nanoporous Metal Layer (NPML):** A multilayered plasmonic structure comprising a nanoporous metallic layer (NPML) and other layers (e.g., a buried dielectric layer) can yield vastly superior sensitivity (e.g., 35500 nm/RIU) and a larger penetration depth compared to standard SPR (7250 nm/RIU) or LRSPR. This is useful for sensing large bioentities like bacteria in water.[92]

○ **Reference Mechanisms:** For multilayered sensors, one resonance dip (e.g., the SP wave excited at the buried-layer-metal interface) can be used as a reference against the sensing dip from the analyte interface. Multichannel planar structures are also used, separating channels containing the analyte from reference channels.[92]

○ **Refractometry (Optical Clock):** Highly precise SPR sensing can be achieved by integrating the plasmonic sensor into a coherent optical resonator, forming a plasmonic optical clock. This method measures the frequency difference (beat signal, BN) between the TE mode (not affected by SPP) and the TM mode (coupled to SPP). This frequency-based measurement is intrinsically immune from amplitude noise and thermal-mechanical noise, leading to high performance even for detecting small refractive index changes.[90]

## B. Localized Surface Plasmon Resonance (LSPR)

LSPR biosensing uses metallic nanoparticles and nanostructures as transduction platforms, relying on the extreme sensitivity of the localized plasmon resonance to the refractive index changes in the local environment.[33], [96], [112], [133]

1. **Sensing Mechanism:** When a target molecule binds to a functionalized nanostructure (e.g., ligands on gold nanostructures), it causes a perturbation in the local index of refraction, which is manifested as a spectral red shift and an increase in intensity of the LSPR.[96], [112]

2. **Detection and Imaging:** LSPR detection is typically performed label-free.[33], [96], [112], [133]

   ○ **Wavelength Shift Measurement:** Detection involves directing light onto the nanotransducers and detecting the reflected light (e.g., with a spectrophotometer). A phase shift or spectral peak wavelength shift (red or blue shift) indicates binding.[33], [113]

   ○ **LSPR Imaging (LSPRi):** High-throughput detection is achieved using LSPR imaging chips, enabling real-time imaging of secreted protein concentrations over time and spatial distance from living cells. This allows for the spatio-temporal mapping of extracellular concentration data.[112]

3. **High Sensitivity and Specificity:**
   ○ **Artificial Antibodies (MIPs):** To overcome the limitations of natural antibodies (cost, stability), synthetic Molecularly Imprinted Polymers (MIPs) are used. These MIPs contain recognition cavities substantially complementary to the target molecule, which are formed on plasmonic nanotransducers (e.g., hollow nanostructure cores like gold nanocages). Detection of the binding complex is then measured using LSPR (or SERS).[33]

◦ **Single-Molecule Resolution (Conformational Sensing):** High sensitivity is achieved by detecting the change in molecular conformation of a binding molecule (Conformational Molecule, CMM) when it binds an analyte. For instance, using a plasmonic dimer (e.g., gold nanorod protrusion and a gold nanosphere target) connected by a CMM (like an aptamer), analyte binding changes the distance/orientation between the two nanostructures, resulting in a measurable shift in the plasmon resonance of the dimer (color change). This technique is sensitive enough for single-molecule detection.[106]

◦ **Exceptional Points (EPs):** Sensors operating at Exceptional Point (EP) singularities in coupled plasmonic nanoresonators demonstrate enhanced sensitivity over other plasmonic sensors (like Raman spectroscopy).[134]

### 3.1.2 Enhancement of Emission and Spectroscopy

Plasmonic effects are used to amplify the signals produced by other optical phenomena, enabling the detection and identification of trace amounts of molecules.

**A. Surface Enhanced Raman Scattering (SERS)**

SERS is a powerful technique where plasmonic nanostructures dramatically amplify the normally weak Raman signal from analyte molecules, providing unique chemical fingerprints.[53], [73], [96]

1. **Mechanism of Enhancement:** The large enhancement (factors as high $as 10^{13} to 10^{15}$ at "hot spots") arises mainly from two mechanisms:[38], [52], [58], [103]

   ◦ **Electromagnetic Enhancement:** The local electromagnetic field is amplified ($up to 10^6$-to $10^7 - fold, or up to 10^{15} - fold at hot spots$) by surface plasmons. This amplifies the Raman emission intensity, which is proportional to the square of the applied field at the molecule.[37], [38], [52], [58], [101], [103]

   ◦ **Chemical Enhancement:** Associated with direct energy transfer between the molecule and the metal surface.[38], [103]

1. **Detection and Specificity:**
   ◦ **Identification:** SERS provides distinct spectral differences due to the strong Raman bands. Analysis of SERS spectra can use multivariable statistical means like Principal Component Analysis (PCA) to classify and differentiate individual biological species (analyte and interfering molecules).[33], [123]

   ◦ **Multiplexing:** The highly specific and narrow Raman spectral bands allow multiple assays to be performed simultaneously in a single sample solution, using multiple Raman labels.[33], [66], [103], [123], [135]

   ◦ **DNA Diagnostics (NPCI):** A technique called nano-network plasmonics coupling interference (NPCI) uses SERS signal intensity changes. Target DNA strands interfere with the plasmonic coupling effect between LNA-NPs and probe-NPs by competing for binding. This interference results in a quenched SERS signal whose intensity decrease is quantitative to the target DNA concentration (e.g., minimum concentration detected around 100 pM).[103]

   ◦ **In Vivo/Remote Sensing:** SERS diagnostics can be expanded to the field using hand-held Raman spectrometers and microfluidics. Sensors (e.g., "nano-bulb" SERS substrates) can be fabricated on the tip of an optical fiber for in-vivo, endoscopic measurements.[135]

## B. Plasmon-Enhanced Fluorescence and Bioluminescence

Plasmons can enhance other light emission phenomena, improving signal-to-noise ratios (SNR) for diagnostics.

1. **Metal-Enhanced Bioluminescence (MEBL):** Chemically induced electronic excited states (bioluminescence species) couple to surface plasmons, producing emission intensities from about 5 to about $1000\,fold$ greater compared to a control sample. This approach is significant for optically amplifying bioluminescence based clinical assays, potentially increasing analyte/biospecies detectability.[19], [111]

2. **Near-Infrared Fluorescence Enhancement (NIR-FE):** Plasmonic gold substrates (discontinuous films forming gold "islands") can afford NIR-FE of up to 100-fold or greater.[66]

   ◦ **Mechanism:** Enhancement results from amplified excitation electric fields in nanogaps and increased radiative decay of excited states due to resonant coupling between plasmonic modes and fluorescent emission dipoles.[66]
   
           ◦ High Sensitivity: This enhancement improves the signal-to-noise ratio over standard protein microarrays (e.g., $\sim 100-fold\ for\ IR800-$

*labeled assays*), achieving detection limits as low as 5 fM for cancer biomarkers like CEA, with a broad dynamic range over six orders of magnitude. This dramatically overcomes the detection limits of traditional methods like ELISA for low-level analytes such as cytokines.[66]

### 3.1.3 Multi-Modal and Advanced Detection Schemes

Plasmonics enables complex and integrated detection systems suitable for demanding clinical applications.

1. **Combined Concentration and Sensing:** Plasmonic substrates are used in systems that first trap/concentrate particles (e.g., biological cells, DNA molecules) using plasmon-induced thermal gradients/hotspots, and then immediately use the same plasmonic components as a sensing substrate for detection. The plasmonic nanostructures can be functionalized with antibodies or aptamers to selectively bind the concentrated target particles.[62], [130]

2. **Exosome Detection:** Nano-plasmonic sensors based on extraordinary optical transmission through predefined patterned nanoapertures in a plasmonic film are used to capture and quantify exosomes. The binding of exosomes to the sensor causes a significant change in the optical signal used to indicate their presence. This method has demonstrated superior sensitivity (e.g., $\sim 10^2$-fold higher than ELISA) for quantitative detection of specific exosomes (e.g., using markers like EpCAM or CD24 for ovarian cancer).[1]

3. Real-Time Cell Activity Monitoring: Plasmonic sensing platforms can extract and quantify the ionic activities of non-voltage gated and voltage-gated ion channels of live excitable cells (e.g., cardiac cells). The sensitivity of plasmons to charge distribution (within 200 nm) makes them suitable for sensing biologically driven charge variations. This non-invasive approach inflicts near-zero disruptions on the cellular micro-environment compared to other techniques like patch clamping.[115]

4. **Multiplexed Diagnostics (Point-of-Care):** Plasmonic devices facilitate portable, high-throughput systems for use in diagnostics, pharmaceutical research, and point-of-care testing (POCT).[132], [133]

   ◦ **Microfluidic Integration:** Microfluidic chips integrate plasmonic sensing (e.g., LSPR or SPR) for the simultaneous detection of multiple analytes like proteins and nucleic acids, using small sample volumes.[133], [135], [136]

   ◦ **Scanning SPR:** Miniaturized SPR devices using rotatable micromirrors (MEMS) allow the device to be small enough to be integrated into devices like cell phones or tablets. The device can scan multi-analyte test strips (analyte areas equipped with different linkers) two-dimensionally, enabling multiplex testing.[132]

## 3.2 Plasmonic Devices for Biosensing and Diagnostics

Plasmonic devices and assays leverage fundamental plasmonic effects to achieve highly sensitive, label-free, and portable biosensing and diagnostic platforms, often reaching single-molecule or femtomolar detection limits. These innovations are tailored for applications ranging from clinical diagnostics to point-of-care testing (POCT). This document categorizes the key devices and assay formats into three primary groups: resonance-based devices (SPR and LSPR), amplified emission and molecular sensing assays, and advanced and integrated diagnostic systems.

### 3.2.1 Resonance-Based Devices and Assays (SPR and LSPR)

These devices detect analytes by monitoring shifts in plasmon resonance caused by changes in the local refractive index upon molecular binding.[92]

**A. Surface Plasmon Resonance (SPR) Systems**

SPR systems generally rely on coupling light into propagating surface plasmon polaritons (SPPs) on a thin metal film.[96], [97]

- **Kretschmann Configuration Systems:** These systems are fundamental to most commercial SPR sensors. They involve coupling excitation light through a high-refractive-index prism adjacent to a thin metal layer, and monitoring the minimum reflected intensity (the SPR dip) which shifts when biological interactions occur on the outer surface. The measured shift in the incidence angle or resonant wavelength indicates the presence or quantity of biological or chemical entities.[92], [97]

- **Multilayered Plasmonic Structure Sensors:** These enhanced SPR sensors incorporate a nanoporous metallic layer (NPML). This structure provides superior sensitivity and a larger penetration depth compared to standard SPR, making them useful for detecting large bioentities like bacteria. They can use one resonance dip as a reference against the sensing dip to minimize noise.[92]

- **SPR Optical Clock/Interferometric Systems:** These high-precision SPR devices integrate the sensor into a coherent optical resonator to measure the frequency difference (beat signal) between coupled and non-coupled optical modes. This technique provides high performance, often immune from amplitude noise and thermal-mechanical noise, for detecting small refractive index changes.[90]

- **Grating-Coupled SPR (GCSPR) Devices:** These devices replace the costly prism with a **diffraction grating** patterned on the metal surface to couple light into the plasmons. This configuration simplifies the optical pathway and allows for the manufacture of larger-scale SPR-active surfaces.[92], [96], [97]

- **Micromirror-Based SPR Sensor Devices:** These portable devices use a tiltable or rotatable micromirror to vary the angle of incidence. This allows the metallic film (often a test strip) to be scanned two-dimensionally. The test strip can be equipped with various linkers to test for different substances (multiplexing) in a format similar to a conventional point-of-care test (POCT), like a blood sugar measuring unit.[132]

## B. Localized Surface Plasmon Resonance (LSPR) Systems

LSPR utilizes metallic nanostructures (nanoparticles or nanoarrays) as sensing elements.

- **Nanoparticle-Based LSPR Biosensors:** These employ the strong UV-Vis absorption band of metal nanoparticles, detecting analyte binding by observing a spectral peak wavelength shift (color change) or intensity change in the reflected light. They are amenable to miniaturized devices because the sensing elements are small, making them suitable for microchip applications and POCT.[33], [92], [113], [133]

- **LSPR Microfluidic Devices and Spot Plates:** Plasmonic nanoparticles arrays, fabricated using methods like Hole Mask Colloidal Lithography combined with photolithography, are integrated into microfluidic channels or multiwell plates (spot/microwell plates).[133]

  ◦ **Microfluidic Channel Devices:** Used for simultaneously assaying two or more samples for a target analyte. These platforms measure the kinetics of multiple analytes by flowing solutions through small channels.[133]

  ◦ **LSPR Spot Plates:** Typically have multiple sample wells (e.g., 6, 24, 96, etc.) with nanoparticle arrays functionalized for different target analytes, allowing multiplexed assays on a single sample. These devices can be read by a standard plate reader or a simple UV-Vis spectrometer.[133]

- **Molecularly Imprinted Plasmonic Nanotransducers (Artificial Antibodies):** These devices replace natural antibodies, which are expensive and unstable, with Molecularly Imprinted Polymers (MIPs) formed onto hollow nanostructure cores (e.g., gold nanocages). The MIPs contain recognition cavities complementary to the target molecule.

Detection of the resulting complex is performed via LSPR or SERS. The ability to reuse the nanotransducers at least about 5 times is noted.[33]

## 3.2.2 Amplified Emission and Molecular Sensing Assays

These methods rely on plasmonic structures to enhance the optical signal from the analyte or a label, enabling trace detection.

### A. Surface Enhanced Raman Scattering (SERS) Devices

SERS substrates amplify the Raman signal by up to $10^{15}$ −fold at "hot spots," providing distinct chemical fingerprints for molecular identification.[37], [38], [103]

- **Porous SERS Analytical Devices (Filter SERS):** These devices incorporate plasmonic nanoparticles (e.g., silver or gold) embedded in a porous substrate (e.g., nylon, PVDF, or paper). The Filter SERS technique leverages the porous nature to trap and concentrate the analyte from large sample volumes, significantly improving the limit of detection for trace chemicals like pesticides (malathion) or food contaminants (melamine). Kits based on this method include plasmonic nanoparticle solutions and a porous membrane configured to trap them to form the SERS sensing region.[137]

- **S-SERS (Stamping SERS):** This technique uses a SERS substrate, such as a nanoporous gold disk (NPGD), which is stamped onto a surface containing the target molecule(s) (e.g., deposited on a PDMS carrier substrate), and SERS measurements are taken directly from the molecules compressed between the surfaces. This enables label-free, multiplexed molecular sensing and imaging.[138]

- **Nanopillar/Nano-bulb SERS Substrates:** SERS nanostructures, like the "nano-bulb" substrates, are exploited for highly sensitive, multiplexed sensing. For instance, they can be functionalized with antibodies or cDNA to simultaneously detect multiple disease markers, such as IL-8 protein and IL-8 mRNA, from a single saliva sample, moving toward early disease detection devices. SERS substrates can also be fabricated on the tip of an optical fiber for *in-vivo*, endoscopic measurements.[135]

- **Bifunctional Nanotubes/Microtubes:** SERS substrates on silica-coated magnetic microspheres allow for trace detection of various samples, including BTEX, pathogens (viruses, bacteria), antigens, and nucleic acids.[60]

### B. Plasmon-Enhanced Luminescence Assays

Plasmonic structures are used to enhance the typically weak signals from fluorescent or luminescent labels.

- **Metal-Enhanced Bioluminescence (MEBL) Assays:** The invention focuses on using metallic particles (nanostructures, islands, colloids) near bioluminescence-based reactions to enhance the signal. Chemically induced electronic excited states from reactions (e.g., involving luciferase and luciferin, or metabolic changes in bioluminescent bacteria) couple to surface plasmons, producing emission intensities 5 to 1000 fold greater than controls. This amplification is significant for optically amplifying bioluminescence based clinical assays, improving analyte detectability. This technique has been demonstrated using Silver Island Films (SiFs) to detect the bioluminescence signatures of *Escherichia coli* strain TV1061, indicating sample toxicity.[111]

- **Metal-Enhanced Fluorescence (MEF) / Near-Infrared Fluorescence Enhancement (NIR-FE):** MEF/NIR-FE is used in assays, such as microarrays, to boost fluorescence intensity, especially in the near-infrared (NIR) range (600-2000 nm). Plasmonic gold substrates (nanoscopic gold films or isolated island areas) are designed to enhance the excitation field and couple resonantly to emission dipoles, increasing the fluorescence radiative decay. This enhancement (up to \sim 100-fold or greater) improves the signal-to-noise ratio in protein microarrays, allowing for the detection of low-level cancer biomarkers (e.g., CEA) with limits as low as 5 fM}.[66], [139]

- **Microwave-Accelerated Plasmonic Assays (MAMEF/SPCC):** These systems apply low power microwave energy to plasmonic detection systems (MEF or Surface Plasmon-Coupled Chemiluminescence (SPCC)). The microwave energy increases heat in the system and/or increases the reaction kinetics (e.g., binding or hybridization rates). This significantly reduces the time required for forming binding aggregates and increases sensitivity.[19], [116]

### 3.2.3 Advanced and Integrated Diagnostic Systems

Plasmonic technology is often integrated into complex platforms for enhanced functionality, high throughput, and analysis of living cells.

- **Microfluidic Nanoplasmonic Systems:** These devices monolithically integrate micro-scale channels and chambers with arrays of nano-scale plasmonic elements (SPR, LSPR, or SERS). These systems are designed to monitor cell behavior, motility, attachment, viability, and biomolecule interactions *in situ*, label-free, and in real-time, while efficiently controlling the cellular microenvironment.[73], [115]

- **Opto-Thermal Tweezers (OTTs) and SERS Tweezers:** These combined systems use plasmon-enhanced thermophoresis for the on-chip concentrating, sorting, and sensing of particles or analytes. Focused light generates thermal hotspots on plasmonic structures, creating confinement regions that trap particles (e.g., cells or DNA). The plasmonic structures in these trapping regions then serve as the sensing substrate (e.g., functionalized with antibodies or aptamers) to detect the concentrated target.[114], [130]

- **Plasmonic Interferometers (Nano-slit/Groove Arrays):** These nanoscale devices (e.g., two grooves flanking a slit in a silver film) are provided for high-throughput, real-time sensing of chemical and biological analytes [177, 179, 9678004B2]. They can be combined with dye chemistry to alter a molecular target's optical properties for detection of clinically relevant analytes like glucose. They are also proposed for measuring cytokine levels in trauma patients.[96]

- **Single-Molecule Conformational Biosensors:** These specialized LSPR sensors use a plasmonic dimer (e.g., a gold nanorod protrusion coupled with a gold nanosphere target) linked by a Conformational Molecule (CMM), such as an aptamer. Binding of a single target analyte causes the CMM to change conformation (e.g., change distance or orientation), resulting in a measurable shift in the dimer's plasmon resonance (color change). This provides single-molecule resolution and high specificity by allowing discrimination between specific and non-specific interactions.[106]

- **Nanoplasmonic Exosome Detectors:** These highly sensitive, label-free devices are based on extraordinary optical transmission (EOT) through patterned nanoapertures in a plasmonic film. They are designed to capture exosomes via marker-specific capture agents, and the binding causes a significant change in the optical signal (e.g., a shift in peak wavelength or intensity change). This method offers detection sensitivity on the order of $10^2$ −fold higher than ELISA. These systems can be made small and portable.[1]

- **Multimodal Biosensors:** These advanced devices combine plasmonic sensing (via nanodisk antennae) with other transduction mechanisms, such as a nanomechanical resonator (mass sensing) and a field effect transistor (electrical sensing) on the same platform. This simultaneous, synergistic operation in optical, electrical, and mechanical modes allows for the determination and differentiation of multiple parameters intrinsic to adsorbed molecules (e.g., mass, binding affinity, surface dissociation constant) on a single device, mitigating the sensitivity-dynamic range trade-off of traditional single-mode sensors.[140]

- **Live Cell Ion Channel Monitors:** These devices use plasmonic sensing platforms (e.g., a gold sensor on a prism) integrated into microfluidic devices. By tracking the interaction of the cell's charge distribution (ionic activities of ion channels) with the plasmon's electron density cloud, the system can **extract, correlate, and quantify action potentials** and properties like conductivity and rhythmicity of **live excitable cells** (e.g., cardiac cells).

This approach is non-invasive, inflicting "near-zero disruptions" on the cellular environment compared to techniques like patch clamping.[115]

- **Nanoplasmonic Imaging for Cell Secretions (LSPRi):** Arrays of gold plasmonic nanostructures are used for real-time imaging of secreted protein concentrations from single cells over time and spatial distance. This method provides spatio-temporal mapping of extracellular concentration data and is useful for analyzing signaling pathways.[112], [141]

- **Exceptional Point (EP) Sensors:** These nanosensors are proposed as a conceptual breakthrough, operating at EP singularities in coupled plasmonic nanoresonators to achieve enhanced sensitivity over other sensors, such as those involving Raman spectroscopy. They are designed to be compact and compatible with biologically relevant substances for genomics, virology, and proteomics.[134]

- **Integrated Analytical Systems/Point-of-Care Devices:** Many plasmonic devices are designed to be portable and integrated with user electronics. Examples include mobile molecular diagnostics systems configured for resource-limited settings, where a smartphone can be used for data image capture, analysis, storage, and wireless transmission.[136]

## 3.3 Plasmonic Biosensing Targets: Biomarkers, Pathogens, and Chemicals

Plasmonic biosensing systems target a diverse array of analytes for diagnostics, focusing on biological and chemical molecules associated with disease, metabolism, and environmental toxins. These systems leverage the high sensitivity and specificity of localized surface plasmon resonance (LSPR), surface plasmon resonance (SPR), and surface-enhanced Raman scattering (SERS) to detect and quantify analytes. This document categorizes the target analytes into three primary groups: cancer and disease-related biomarkers, general analytes in complex biological samples, and environmental and chemical targets.

### 3.3.1 Cancer and Disease-Related Biomarkers

A significant focus of plasmonic biosensing research detailed in the sources is the detection of markers for cancer and other serious diseases.

**A. Nucleic Acids Related to Cancer**

Plasmonic sensors are designed to detect and quantify nucleic acids, often used as biomarkers for cancer diagnosis or genetic analysis:

- **Free DNA (ctDNA) / Genomic DNA:** Quantifying free DNA, particularly circulating tumor DNA (ctDNA) derived from cancer cells in body fluids like blood, is considered a method for diagnosing cancer. It is noted that the amount of free DNA from cancer cells is significantly higher in cancer patients than in healthy subjects.[142]

- **Nucleosomes and Chromatin:** A primary focus of specialized plasmonic chips is the selective absorption and imaging of cancer-related substances in serum, specifically histone wound around with DNA (nucleosomes) and chromatin (fiber) where histones form a string-like structure. Thes nucleosomes/chromatin increase dramatically (up to 100 times) with the progression of cancer, unlike other positively charged proteins like globulin (which increases up to 2 times at most). The detection method relies on the positively charged protein molecules in these cancer-related substances being captured by the negatively polarized meso-crystals on the plasmonic chip.[142]

- **Specific DNA Sequences:** Plasmonic probes are used in DNA diagnostics via Plasmonics Coupling Interference (NPCI). Target analytes detected include:[103]

  ◦ **Nucleic Acids (General):** DNA, RNA, single-stranded, double-stranded, or triple-stranded, and chemical modifications thereof.[53]

  ◦ **Specific DNA Targets:** The NPCI method detects target DNA strands.[103]

  ◦ **MicroRNA (miRNA) and mRNA:** Specific miRNA sequences can be detected as a screening tool for cancer diagnosis. Plasmonic chips are also capable of detecting and quantifying IL-8 coding mRNA (a pre-cancer marker). Intravesicular markers (markers inside exosomes) include mRNA, microRNA, lncRNA, and DNA.[1], [103], [135]

  ◦ **Single Nucleotide Polymorphisms (SNPs):** Detection methods are demonstrated for identifying SNPs (e.g., $ERBB2\ Ile654Val\ target\ DNA$).[103]

  ◦ **Peptide Nucleic Acid-DNA Hybrids:** Label-free detection of peptide nucleic acid-DNA hybridization has been demonstrated using LSPR biosensors.[112]

**B. Proteins and Peptides (Cancer and General Disease Markers)**

Proteins, peptides, and related structures are common targets, often detected using antibody/antigen binding partners.[37], [101], [113], [125]

- **Tumor Markers/Antigens:** Proteins and peptides that serve as tumor markers or cancer-related antigens are major targets. Examples include:[33], [113], [142]

  ◦ **NGAL (Neutrophil Gelatinase-Associated Lipocalin):** A urinary protein biomarker detected in urine for kidney injury.[33]

  ◦ **CEA (Carcinoembryonic Antigen):** A glycoprotein and cancer biomarker.[66]

  ◦ **PSA (Prostate-Specific Antigen), AFP, CA125, CA15-3, CA19-9, MUC1, GM3, GD2, ERBB2, NY-ESO-1:** Identified as disease-related antigens or cancer-related antigens/markers.[1], [113]

  ◦ **EGFR (Epidermal Growth Factor Receptor) and Her2:** Specific tumor markers targeted by antibodies conjugated to plasmonic agents for therapy and diagnostics (e.g., anti-EGFR antibodies for oral and cervical cancer cells; anti-Her2 antibodies for breast cancer cells).[37], [52], [58], [101]

- **Extracellular Proteins/Secretions (Live Cell Monitoring):** Plasmonic sensors (LSPRi) are used to measure the extracellular concentrations of secreted proteins in real-time from single living cells. The technique is applicable to numerous analytes and cell types, helping resolve concentrations and monitor burst-like secretions.[112], [141]

- **Cytokines and Inflammatory Markers:** Plasmonic biosensors target cytokines (a class of low molecular weight nonantibody proteins). Specific examples include:[66]

  ◦ **IL-8 protein:** A pre-cancer marker detected in saliva.[135]
  ◦ $IL-1\beta, IL-4, IL-6, IFN-\gamma$, **and TNF:** Targets analyzed in multiplexed cytokine assays.[66], [135]

  ◦ **C-reactive protein (CRP) and other inflammatory markers (MRP14, MRP8, 25F9):** Targets for disease diagnosis.[113]

- **Cardiovascular Disease Markers:** Targets include troponin, C-reactive protein, brain natriuretic peptide, CKMB, and fatty acid binding protein.[113], [125]

- **Metabolic and Hormonal Markers:** Targets include hormones, thyroid stimulating hormone, thyroxine, leptin, and insulin. Glucose sensing has been described using nanogold plasmon resonance-based methods.[106], [113], [123], [125]

- **Autoimmune Disease Markers:** Plasmonic assays target auto-antibodies.[113], [125]

## C. Large Cellular Structures and Pathogens

- **Exosomes and Nanovesicles:** Specialized nano-plasmonic sensors are designed for highly sensitive, label-free detection and quantitation of exosomes (membrane-bound phospholipid nanovesicles, 50-200 nm in diameter). Exosomes are crucial targets because they carry molecular information reflecting the parent tumor.[1]

  ◦ **Exosome Markers:** Plasmonic devices target specific exosome surface markers (extravesicular) or internal components (intravesicular). Examples include EpCAM, CD24, CA19-9, CA-125, HER2, CD63, CD9, CD81, TSG101, MUC18, EGFR, HSP90, HSP70, and many others. Intravesicular markers can be protein, lipids, small molecules, mRNA, microRNA, lncRNA, and DNA.[1]

- **Bacteria and Viruses:** Plasmonic systems are used to detect and monitor pathogens.

  ◦ **Bacteria:** Specific targets include *Escherichia coli* strain TV1061 (used to demonstrate metal-enhanced bioluminescence amplification), *Bordetella pertussis* (Bp, whooping cough) and its secretions, such as tracheal cytotoxin (TCT), filamentous hemagglutinin (FHA), and Pertussis Toxin (PTx). Plasmonic-magnetic bifunctional nanotubes are also suggested for detecting bacteria.[60], [116], [135]

  ◦ **Viruses:** Examples include Zika-virus (ZIKV) envelope protein and its specific antibody (using toroidal resonance sensors). General viral antigens (e.g., feline leukemia virus, canine parvovirus, hepatitis a, b, c virus, HIV, human papilloma virus, Epstein Barr virus, rabies virus) are listed as potential targets.[49], [125]

  ◦ **Other Pathogens:** Plasmonic-magnetic nanotubes are intended for detecting spores, fungi, and parasitic antigens (*Giardia lamblia, Plasmodium falciparum, Trypanosoma brucei, canine heartworm*).[60], [125]

## 3.3.2   General Analytes in Complex Biological Samples

Plasmonic biosensing methods are capable of analyzing a broad scope of molecules present in biological fluids (whole blood, plasma, serum, urine, saliva, cerebrospinal fluid, sweat).[33]

- **General Molecular Classes:** Target analytes universally include cells, proteins, peptides, nucleic acids (RNA, DNA, mRNA, miRNA), hormones, glycoproteins, polysaccharides, toxins, virus particles, drug molecules, haptens, and chemicals.[33], [113], [125]

- **General Biological Matter:** The non-invasive sensors (like those tracking ion channel activities) analyze the electrical/mechanical activity of live excitable cells (e.g., cardiac cells).[115]

- **Small Molecules and Metabolites:** Electrolytes, metabolites, small molecules, and drugs are identified as targets.[106]

- **Interfering Molecules:** Sensing platforms must contend with common interfering molecules in physiological samples, such as Bovine Serum Albumin (BSA). Note that globlin, DNA left from normal cells, DNA dissociated from histones by acetylation, and albumin are generally not absorbed by certain cancer diagnostic chips due to charge selectivity.[33], [142]

### 3.3.3 Environmental and Chemical Targets

While focused on biosensing, the underlying technology is applicable to general chemical detection:

- **Environmental Toxins and Chemicals:** SERS substrates on magnetic microspheres are designed to detect trace samples including BTEX (benzene, toluene, ethylbenzene, and xylenes), chlorinated solvents, TNT, nerve agents, and blister agents.[60]

- **Inorganic/General Chemicals:** Targets also include metal ions and anions. Gas phase chemical detection is proposed using plasmonic structures coupled with adsorption materials for detecting target gases.[60], [123]

## 3.4 Plasmonic Structures for Energy and Optical Applications

Plasmonic structures exploit unique electromagnetic properties—such as enhanced absorption, localized field confinement, resonant coupling, and non-radiative energy transfer—to enable high-efficiency, miniaturized, and novel functionalities in energy conversion, solid-state lighting, lasers, and optical communication. This document categorizes these applications into three primary areas: energy conversion (optical to electrical power), emission and amplification for lighting, and optical devices for communication.

### 3.4.1 Energy Conversion: Optical to Electrical Power (Photovoltaics and Thermophotovoltaics)

Plasmonic structures are utilized extensively to improve the efficiency of solar energy harvesting and conversion into electrical power, particularly in thin-film or organic devices where enhanced light absorption is critical.[143], [144]

**A. Plasmoelectric Conversion (Direct Optical-to-DC Electrical Power)**

A novel conversion method involves the "plasmoelectric effect," which directly converts absorbed optical power into direct current (DC) electrical power using plasmonic nanostructures.[102]

- **Mechanism:** The plasmoelectric effect relies on resonant optical absorption in plasmonic nanostructures (e.g., gold (Au) and silver (Ag) objects of nanoscale size). The process exploits the dependence of the plasmon resonance frequency ($\omega_p$) on the electron density (n) of the nanostructure. When irradiated, changes in the internal charge density induce charge transport between electrically coupled nanostructures, generating a plasmoelectric potential and current.[102]

- **Efficiency Improvement:** This approach addresses the problem of low optical-to-electrical power conversion efficiency found in typical schemes, which are limited by the very short excited state lifetime (usually less than 10 fs) of excited electrons in metals. The plasmoelectric device efficiency is not limited by this short excited state lifetime.[102]

- **All-Conductor Circuitry:** Power conversion results from optically generated heat affecting resonance shifts, meaning no additional semiconductor or insulating components (like Schottky barriers or P-N junctions) are needed to rectify the excited electrons. This allows for an all-metal, or all-conductor, circuit to convert optical power to electrical power.[102]

- **Device Design:** Plasmoelectric devices typically consist of first and second electrically coupled plasmonic nanostructures, potentially arranged in arrays for broadband power conversion. They maximize efficiency by using strategic device geometry and customized incident radiation profiles. Charge transfer can be induced when nanostructures receive resonant irradiation at a value *between* their respective resonance values, or when they receive off-resonant irradiation (outside the resonance range) to induce charge transfer by altering internal charge densities.[102]

## B. Solar Thermophotovoltaics (S-TPV)

Plasmonic structures are crucial for improving the efficiency of S-TPV systems by controlling absorption and emission spectra.[145], [146]

- **Integrated Spectrally-Selective Plasmonic Absorber/Emitter (ISSAE):** A novel thin-film ISSAE device functions as three components simultaneously:

- An efficient broad-band absorber of solar radiation (e.g., in the $250\ nm < \lambda < 1\ \mu m\ range$).[145]

- An efficient heat insulator by being an extremely poor absorber/emitter of far-infrared radiation (e.g., for $\lambda > 3.5\ \mu m$).[145]

- A spectrally selective infrared emitter that can be tuned to emit photons in a specific range (e.g., $1.5\ \mu m < \lambda < 2.5\ \mu m$), thereby matching the bandgap of the targeted photovoltaic cell.[145]

- **Enhanced Efficiency:** This spectral control is essential because it enables high conversion efficiency by ensuring the emitter supplies IR photons of the "right energy" to the photovoltaic cell. By minimizing infrared absorption, the surface emits much less infrared radiation when hot (Kirchhoff's Law), enhancing the efficiency of day-and-night TPV systems that produce electricity even without sunlight.[145]

- IR Emitters/Sources: Plasmonic structures are also used specifically as IR emitters or sources. Designs, often based on periodic structures created on a surface, can tune the emissivity at particular wavelengths, such as emitting peaks near $4\ \mu m\ and\ 6\ \mu m$. Plasmonic IR emitters can be integrated into MEMS (Microelectromechanical Systems) based devices to isolate heat and reduce power consumption, which is critical for miniaturized IR gas sensors.[146]

## C. Enhanced Photovoltaic Cells (Hot Carrier and Light Trapping)

Plasmonics improves traditional photovoltaic and photodetector devices by enhancing light interaction within the active layers.[22], [102]

- **Light Trapping and Absorption:** Plasmonic nanostructures, such as those made of gold (Au) and silver (Ag), are used as concentration schemes for photovoltaics. They trap and manipulate incident light, leading to the excitation of surface plasmon resonances and high absorption. This has yielded increased photocurrents in solar cells and resulted in enhancement of the photovoltaic conversion efficiency of devices.[2], [22], [102], [144]

- **Organic Photovoltaic (OPV) Cells:** Plasmonic nanocavity arrays can significantly increase the power conversion efficiency of organic photovoltaic cells (OPV), providing a substantial fold increase compared to similar devices without patterned metallic anodes. This involves creating an array of plasmonic nanocavities within the active semiconductor region, which captures and channels surface plasmon waves to increase optical field intensity.[144]

- **Plasmon-Induced Hot Carrier Devices:** These devices use plasmon resonance to generate hot electrons that overcome a potential barrier (Schottky barrier) between the plasmonic material (e.g., silver) and a semiconductor (e.g., silicon). This mechanism allows conversion of light with photon energy below the semiconductor band gap (e.g., IR light) into a useful photovoltage, thereby increasing the overall energy conversion efficiency of solar cells.[107]

- **Rectenna Devices:** Plasmonic rectennas (rectifying antennas) are being developed for harvesting solar energy, utilizing enhanced fields and electron emission.[147]

## D. Plasmon-Enhanced Photocatalysis (Chemical Conversion)

Plasmonics provides a platform to efficiently convert light energy into chemical work by driving desirable chemical reactions.[5], [97]

- **Antenna-Reactor Effect:** Plasmonic materials act as an antenna, generating "hot carriers" through plasmon decay, which then transfer to an electronically conductive reactive catalytic surface (reactor) to drive chemical reactions. The plasmonic material also converts light energy into local heat (thermal coupling), which can thermally drive reactions on the associated reactive component.[5], [21]

- **Efficiency:** Full-spectrum coupling with optimized plasmonic gratings and photocatalysts is expected to yield profound improvements in catalytic efficiency. Experiments have demonstrated SPR-enhancement of photocurrent generation (e.g., $4.6 \times enhancement$) in systems optimized for photocatalysis.[97]

- Applications: Plasmon-enhanced photocatalytic systems are proposed for large- and small-scale chemical conversion, including water splitting and $H_2$ utilization collection, self-cleaning materials, and self-sterilizing materials.[97]

### 3.4.2 Emission, Amplification, and Solid-State Lighting

Plasmonics is used to control light emission direction, efficiency, speed, and color in various light sources.

A. Plasmonic Lasers and Amplifiers

Plasmonic structures enable the creation of miniaturized light sources with fast modulation and subwavelength confinement.

- **Plasmonic Nanolasers/Spasers:** These lasers exploit surface plasmon polaritons (SPPs) to achieve deep subwavelength scale operation. They can be electrically pumped and configured with a coaxial plasmonic cavity, comprising a peripheral ring structure and a central core, separated by a semiconductor gain structure (e.g., one or more radial quantum wells).[41], [46], [75], [148]

- **Enhanced Emission Rate:** Plasmonic LEDs and lasers exploit the Purcell effect, positioning a quantum well close to a metal surface to increase the electron-hole pair recombination rate via plasmons. This enhances the spontaneous emission rate and decreases the delay between current change and light emission, allowing for modulation speeds of 10 GHz or faster.[149]

- **On-Chip Integration (EPIC):** Plasmonic lasers are sought for on-chip communication and integrated applications. A plasmonic light source can be formed on a silicon substrate containing CMOS devices, providing an integrated on-chip light source. These systems (EPICs - Electronic Plasmonic-Integrated Circuits) utilize bidirectional conversion between electrical signals and SPPs using materials like Transition Metal Dichalcogenides (TMDs) (e.g., $WS_2, MoS_2$) that can both emit and absorb light.[9], [148], [150]

- **Optical Amplification:** Plasmonic structures act as nanoscale and microscale amplifiers and lasers. Whispering Gallery Mode (WGM) resonators can employ plasmonic nanoantennas to efficiently deliver optical radiation into the resonator.[67]

### B. Organic Light Emitting Diodes (OLEDs) and Solid-State Illumination

Plasmonics is applied to OLEDs to overcome efficiency limitations by managing energy typically lost into non-radiative modes.[151], [152]

- **Energy Transfer and Outcoupling:** A plasmonic OLED incorporates an enhancement layer (plasmonic material) that non-radiatively couples to the organic emissive material, transferring the excited state energy (a majority of it, typically >50%) to a non-radiative surface plasmon polariton (SPP) mode.[151], [152]

- **Converting Plasmon Energy to Light:** To prevent this energy from being lost as heat, the device includes an outcoupling layer (e.g., gratings, nanopatch antennas, or through-hole arrays). This layer scatters the energy from the SPP mode into photons for emission (e.g., top or bottom emission), improving the overall efficiency.[122], [151], [152]

- **Device Enhancement:** The enhancement layer modifies the electromagnetic environment, leading to improved efficiency, increased lifetime (e.g., for blue-emitting devices), and altered emission directionality.[122], [151], [153]

- Light Enhancement for Emitters: The enhancement of spontaneous emission is achieved by coupling light emitters (e.g., colloidal nanocrystals like CdSe/ZnS Quantum dots) to spatially controlled surface plasmons, resulting in a large enhancement of fluorescence efficiency.[11], [41], [154]

## C. Wavelength Conversion and Harmonic Generation

Plasmonics is used to generate light at different frequencies and achieve wavelength conversion.

- **Up-Conversion (Harmonic Generation):** A plasmonic up-converter apparatus uses an array of nanofeatures (e.g., gold, silver, platinum nanoparticles) to produce an emission spectrum containing peaks at second harmonic and third harmonic frequencies ($2\omega, 3\omega$) that are greater than the excitation frequency ($\omega$). This conversion can turn non-visible infrared (IR) electromagnetic radiation into visible or ultraviolet light. This generated high-frequency light can then be used to optically pump a lasing medium.[155]

- **ENZ Materials for Nonlinear Optics:** Gate-tunable Epsilon-Near-Zero (ENZ) materials, such as Transparent Conducting Oxides (TCOs), can electrically and nonlinearly optically tune plasmonic nanocircuits for advanced light manipulation, including enhancement of nonlinear-optical effects. Third harmonic (TH) generation is a process where ENZ materials play an important role, offering larger field enhancements due to deeply sub-wavelength modal volumes, which is crucial for nanoscale optoelectronics.[8], [98]

- **Plasmonics-Enhanced Energy Upconversion (PE-UC):** This technique involves an energy cascade where an applied initiation energy is absorbed, intensified, or modified by a plasmonics agent into energy capable of being upconverted by an energy modulation agent (nanoparticle) into an energy that induces a predetermined change. This is used in therapeutic applications to amplify phototherapy effects.[37], [38], [58]

### 3.4.3 Optical Devices and Communication Components

Plasmonics enables the miniaturization and acceleration of components for data transmission and processing.

- **Optical Modulators and Switches:** Plasmonic confinement allows for significantly greater integration density and lower power consumption in optoelectronic circuits.[12]
  ◦ **Phase Change Materials (PCMs):** Plasmonic modulators can integrate solid-state phase change materials (e.g., $VO_2$) that, when thermally or electrically triggered, switch state to control the propagation of SPPs.[12]

  ◦ **Voltage-Controlled Modulation:** Devices can use voltage-tunable refractive index material (e.g., liquid crystal or TCOs) integrated with plasmonic structures to actively tune the optical properties for applications like dynamic color filters and optical switching.[8], [12], [151], [156], [157], [158]

- **Plasmonic Waveguides and Interconnects:** Plasmonics facilitates subwavelength confinement of light in metallic nanostructures, allowing circuits to be miniaturized to the nanometer scale, matching electronics. Waveguides (e.g., Ag nanowires or slot waveguides) are integrated into Electronic Photonic (Plasmonic) Integrated Circuits (EPICs) to transmit signals between emitters and photodetectors.[2], [9], [12], [159]

- **Terahertz (THz) Generation and Detection:** Plasmonic structures, particularly plasmonic contact electrode gratings, are essential for increasing the optical-to-terahertz conversion efficiency of photomixers, and enabling the generation and detection of high-power, ultrafast THz radiation.[63], [160], [161], [162]

- **Optical Beam Steering and Phased Arrays:** Plasmonic metasurfaces and reconfigurable antenna apertures populated with adjustable plasmonic resonant waveguides are used for optical beam steering in applications like LiDAR and free-space optical communication.[89]

- **Plasmonic Lenses and Focusing:** Plasmonic light collectors and lenses exploit the interaction between incident light and plasmons to redirect and focus light through subwavelength openings. This is critical for achieving high resolution in imaging and efficiently coupling light into transmission cables with less signal loss.[70], [163]

## 3.5 Plasmonics in Data Storage and Communication

Plasmonic structures enable next-generation data storage and communication by overcoming the diffraction limit of light, increasing data density, enhancing component speed, and facilitating nanoscale integration of optics with electronics. This document categorizes these applications into two primary areas: data storage, focusing on heat-assisted magnetic recording (HAMR) and optical storage, and communication and signal processing, encompassing integrated circuits, modulation, and advanced communication systems.

### 3.5.1 Data Storage Applications

Plasmonic devices are integral to increasing storage density and efficiency, particularly through Heat-Assisted Magnetic Recording (HAMR) and novel optical storage techniques.

**A. Heat-Assisted Magnetic Recording (HAMR)**

HAMR is a leading storage technology that utilizes plasmonic near-field energy to overcome thermal stability limits in magnetic media, thereby increasing recording density.[3], [4]

- **Near-Field Transducers (NFTs):** The core plasmonic component is the Near-Field Transducer (NFT), which is illuminated by a light source (e.g., an infrared or near-infrared laser diode). The NFT couples light into free electron plasma to create a plasmon wave, resulting in a highly concentrated electromagnetic field (near-field light).[3], [4], [84]

- **Localized Heating for Writing:** The near-field light is focused to a diffraction-limited spot. The concentrated optical fields locally heat a small region of the recording medium (such as an FePt ) for long-term storage.[3], [4], [84], [164], [165]

- **Increased Recording Density:** The use of the focused plasmonic field allows for extreme light concentration (equal to or less than about 10 nm), forming a light spot smaller than the optical diffraction limit. This high resolution increases recording density. The use of plasmonic devices, including a plasmonic underlayer (PUL), efficiently confines optical fields at a nanoscale to locally heat the recording medium.[84], [165]

- **Thermal Management and Reliability:** Plasmonic materials like gold (Au), silver (Ag), and copper (Cu) are commonly used, but they can suffer from high operating temperatures and thermal instability, leading to damage or deformation of the NFT. To address this:[16], [87], [166], [167]

◦ **Mechanically Robust Materials:** New NFTs are disclosed using mechanically robust materials, such as plasmonic ceramics (e.g., $TiNx, ZrNx, HfN_x$). These ceramics offer high thermal stability and performance at the high temperatures ($up to 500 \setminus circ C$) required for HAMR.[3]

◦ **Layered Near-Field Transducers (NFTs):** Multilayered NFT structures (e.g., alternating Au/TiW or Au/Cr layers) enhance mechanical performance and reliability while maintaining acceptable optical coupling efficiency (e.g., 80% or more compared to solid Au).[168]

◦ **Plasmonic Underlayers (PULs):** A Patterned Plasmonic Underlayer (PUL), located below the recording layer, helps avoid lateral thermal bloom by regulating optical fields and confining heating. Optical-coupling multilayers of alternating plasmonic and non-plasmonic materials (e.g., RuAl/Au or RuAl/Rh) are used as part of the thermal management stack to improve the ratio of Thermal Gradient/Laser Power ($TG/LP$), which prolongs NFT life and improves recording sharpness.[84], [164], [169]

• **Circularly Polarized Fields:** Optical elements containing ring-type or discrete ring-type structures generate a circularly polarized plasmonic field. This field improves writability and recording performance on high coercivity media (like FePt in an $L1_0$ structure).[165]

## B. Novel Optical Data Recording and Storage

Plasmonic nanostructures enable extremely high-density, multi-dimensional optical storage methods far surpassing conventional CD storage densities.

• **Frequency-Modulated (FM) Coding and Storage:** Data can be stored by varying the resonance frequency of plasmonic-dielectric nanostructures.[72]

◦ **Mechanism:** The storage medium consists of a flat transparent substrate divided into nanoscale cells (side dimension d on the order of tens of nanometers). Each cell holds a two-layer concentric core-shell nanostructure (e.g., $SiO_2$ core and silver shell). Data is encoded by selecting the ratio of radii and/or the aspect ratio of the layers in each nanostructure, which dictates its unique plasmonic resonant frequency.[72]

◦ **Readout:** Data is read by illuminating the substrate with broadband infrared (IR) or visible light, typically coupled via an evanescent wave, and detecting the scattering amplitude peak using a Near-Field Scanning Optical Microscope (NSOM).[72]

- **N-ary Data Storage:** By choosing N different ratios of radii and/or aspect ratios, the system can store N-ary data (binary, trinary, or general N-ary) corresponding to N different plasmonic resonant frequencies.[72]

- **High Density and 3-D Storage:** The nanoscale size of the cells allows for high data density; for instance, a $1\ \mu m \times 1\ \mu m$ area can hold hundreds of cells, compared to only four in a regular compact disk. The concept can be extended to 3-D data storage by stacking several layers of transparent substrates, each with printed nanostructures.[72]

## 3.5.2 Communication and Signal Processing Applications

Plasmonic technology addresses the limitations of purely electronic circuits by integrating high-speed optical data transmission at the nanoscale.

### A. Integrated Plasmonic Circuits and On-Chip Communication

Plasmonics is viewed as a next-generation technology poised to exceed the capabilities of silicon and Moore's law by integrating electronic and photonic circuits on a chip.[9]

- **Electronic Plasmonic-Integrated Circuits (EPICs):** EPICs take advantage of high-speed optical communication and highly integrated fast electronics. Plasmons provide sub-wavelength spatial confinement down to the nanometer scale, which is compatible with electronics for nanoscale optical communication on chip level between transistors.[9], [170]

- **Plasmonic Waveguides and Interconnects:** Plasmonic waveguides, often hybrid structures (e.g., combining metal and a dielectric), are fabricated to utilize resonant plasmonic energy to transmit optical energy or information over a distance.[18], [45], [76], [170]

  ◦ **Advantages:** Plasmonic circuits do not require the surface plasmon to be sustained over a long distance (e.g., centimeters) because the active area of the device can be smaller than the wavelength of light. The shorter plasmon wavelength compared to light allows for smaller devices than currently used in semiconductor arts.[6]

  ◦ **Bidirectional Conversion:** Newly proposed EPIC systems utilize Transition Metal Dichalcogenides (TMDs) (e.g., $WS_2, MoS_2$) integrated with plasmonic waveguides (e.g., silver nanowires) to achieve bidirectional conversion between electrical signals and surface plasmon polaritons (SPPs), enabling a complete integrated circuit functionality (emitter, photodetector, or both). This reduces signal loss and enhances device performance in semiconductor packages.[9], [159]

- **Plasmonic Lasers (Radiation Sources):** Plasmonic quantum well lasers, based on SPPs, are instrumental as candidates for radiation sources in large-scale integrated (LSI) circuits, offering alternatives to pure electric communication aids. These nanolasers provide deep subwavelength scale operation and are sought for on-chip communication and integration applications.[41], [46], [75], [148]

## B. Modulation, Switching, and Logic

Plasmonic devices enable ultrafast, compact modulation of optical signals necessary for high-speed data transmission.

- **Optical Modulators and Switches:** Plasmonic switches and modulators control the amplitude or phase of optical signals at speeds far surpassing traditional systems (which are often limited to microsecond response times).[12], [158]

  ◦ **Phase Change Materials (PCMs):** Plasmonic modulators can incorporate solid-state phase change materials (e.g., $VO_2$) that switch between optical states (e.g., insulator to metallic phase) when electrically, thermally, or optically triggered, providing modulation capability.[12]

  ◦ **Graphene-Based Modulators:** Graphene and Transparent Conducting Oxides (TCOs) are used in plasmonic phase modulators where a control voltage (gating) modulates the propagation speed and phase of the SPP wave.[68], [85]

  ◦ **Switching Circuits and Logic:** A plasmonic circuit can be realized by coupling a planar plasmonic device with a controlling gate structure to a complementary plasmonic device, forming a switching circuit. Combinations of these circuit elements can be used to construct more advanced circuits and complex logical devices, analogous to complementary transistor devices.[6]

- **Photoswitchable Waveguides (Optically Rewritable):** Nanostructured photonic materials demonstrating photoswitchable Rabi splitting can be used as optically rewritable photonic waveguides. These materials include photochromic molecules that undergo reversible isomerization upon UV and green light illumination, dynamically controlling the plasmon-waveguide mode coupling.[14]

## C. Advanced Communication Systems (THz and Beam Steering)

Plasmonics supports communication in difficult spectral ranges and advanced free-space configurations.

- **Terahertz (THz) Communication:** Plasmonic structures are critical for improving applications involving the generation and detection of electromagnetic radiation at terahertz (THz) frequencies. Plasmonic nano-antennas and metamaterials are key components in high-speed THz communication systems.[63], [85], [171]

  ◦ **Ultra Massive MIMO:** Communication systems can include a two-dimensional array of plasmonic nano-antennas, with each antenna supporting an SPP wave corresponding to a signal. The very small size of these arrays allows their integration into various communication devices, facilitating Ultra Massive MIMO (Multiple-Input Multiple-Output) configurations.[171]

  ◦ **Enhanced Conversion:** Plasmonic enhancement improves the efficiency of THz photoconductive switches.[63]

  **Free-Space Optical Communication:** Plasmonic metasurfaces and reconfigurable antenna apertures populated with adjustable plasmonic resonant waveguides are used for optical beam steering. Applications include LiDAR (light detection and ranging) and free-space optical communication (e.g., single-beam and MIMO configurations).[89]

- **Fiber Optic Coupling:** Plasmonic light collectors can be used to form light pipes or lenses for efficiently injecting optical communications into a fiber optic cable. Plasmonic devices can also be integrated into optical fiber tips for advanced light manipulation and communication.[8], [163]

- **Exceptional Point (EP) Applications:** Plasmonic nanostructures operating at Exceptional Point (EP) singularities are proposed for applications in communication, offering high sensitivity and performance benefits.[134]

## 3.6 Plasmonic Structures in Biomedical Applications: Diagnostics and Therapy

Plasmonic structures play a pivotal role in biomedical applications, leveraging plasmonic phenomena for ultrasensitive diagnostics, real-time physiological monitoring, targeted drug delivery, and advanced therapeutics. These applications are categorized into three primary areas: diagnostics and biosensing, therapeutic interventions, and integrated cell analysis systems.

### 3.6.1 Diagnostics and Biosensing (In Vitro and In Vivo)

Plasmonic structures are utilized in devices and assays to detect various target analytes with high sensitivity, often achieving label-free, multiplexed, or single-molecule detection.

**A. Cancer and Disease Biomarker Detection**

Plasmonic sensors are extensively applied to identify and quantify markers associated with major diseases, particularly cancer.

1. **Protein and Cancer Marker Detection:**

   ◦ **Luminescence/Fluorescence Enhancement:** Metal-Enhanced Bioluminescence (MEBL) significantly amplifies the signal (5 to 1000 fold) from bioluminescent clinical assays, enhancing the detectability of biospecies. Plasmonic gold films enable Near-Infrared Fluorescence Enhancement (NIR-FE), improving the sensitivity of protein microarrays for early detection of cancer biomarkers like Carcinoembryonic Antigen (CEA) in mouse serum.[19], [66], [111]

   ◦ **SPR/LSPR Biosensors:** LSPR and SPR systems are used for the detection, identification, and quantification of biomolecules. Assays target specific markers such as autoantibodies implicated in autoimmune diseases and cytokines (e.g., $IL-1\beta, IL-4, IL-6, IFN-\gamma,$ and TNF) for diagnostics. Toroidal plasmonic metasurfaces, particularly compatible with human tissues due to the low energy of Terahertz (THz) radiation, are a reliable platform for immunosensing applications and have been used to detect Zika-virus (ZIKV) envelope protein using a specific ZIKV antibody.[27], [49], [66], [111]

   ◦ **Cancer-Related Substances in Serum:** Specialized plasmonic chips use LSPR to selectively absorb and image cancer-related substances in serum, specifically nucleosomes (histone wound with DNA) and chromatin, whose concentration increases dramatically with cancer progression.[142]

2. **Nucleic Acid Diagnostics:**

   ◦ **SERS-Based Diagnostics:** Plasmonic nanoprobes are utilized in the nano-network plasmonics coupling interference (NPCI) method for nucleic acid diagnostics, detecting single-nucleotide polymorphism (SNP) and microRNA sequences associated with breast cancers.[103]

○ **Infectious Disease Markers:** Nanostructures with high Surface-Enhanced Raman Scattering (SERS) enhancement, like the "nano-bulb" substrates, are proposed for detecting disease markers, such as IL-8 protein and IL-8 coding mRNA (pre-cancer markers) from a single saliva sample. These SERS substrates can be embedded in optical fibers for in-vivo, endoscopic chemical detection within organisms.[135]

○ **Bacterial Pathogenesis Monitoring:** The devices are used to measure the secretion of tracheal cytotoxin (TCT) from *Bordetella pertussis* (whooping cough) colonies *in vitro* or *in situ*. TCT is responsible for epithelial damage in the lungs.[135]

## B. High-Throughput and Point-of-Care (POCT) Systems

Plasmonic systems facilitate the creation of miniaturized, rapid, and high-throughput diagnostic tools suitable for clinical and field use.[33], [96]

- **Microfluidic Integration:** Monolithic integrated nanoplasmonic microfluidic systems are designed for use in cell culture applications, monitoring molecular binding, cell behavior, motility, attachment, and viability *in situ* and in real-time. The devices allow for the measurement of the cellular response to different biochemical cues supplemented in the perfusion media.[73]

- **Plasmonic Biosensors with Artificial Antibodies:** To overcome the cost and poor stability of natural antibodies used in assays like ELISAs, LSPR/SERS biosensors are integrated with Molecularly Imprinted Polymers (MIPs), acting as artificial antibodies, for label-free sensing in disease diagnosis and toxicology testing.[27], [33]

- **PCR Amplification:** Plasmonic nanostructures are used in devices to achieve rapid heating for Polymerase Chain Reaction (PCR) amplification of nucleic acids. This uses plasmonic heating apparatus driven by LEDs for ultrafast results.[39], [172]

- **Exceptional Point (EP) Sensors:** Plasmonic nanostructures operating at Exceptional Point singularities are being developed for applications in sensing and communication, promising enhanced sensitivity compared to standard plasmonic sensors.[134]

## 3.6.2  Therapeutic Interventions and Drug Delivery

Plasmonics provides platforms for non-invasive treatment and precision drug delivery, primarily leveraging photothermal effects.

## A. Plasmonics-Enhanced Phototherapy (PEPST and EPEP)

Plasmonic structures amplify light-based therapies, particularly for treating cell proliferation disorders like cancer.

- **Photothermal Cancer Therapy:** The plasmonic response of nanoparticles is utilized in photothermal cancer therapy. Noble metal nanoparticles (e.g., gold) are used as drug delivery platforms that preferentially accumulate within tumor microenvironments via the Enhanced Permeability and Retention (EPR) effect.[37], [52]

- **PEPST Mechanism:** Plasmonics Enhanced PhotoSpectral Therapy (PEPST) probes consist of photo-active (PA) drug molecules (e.g., aminolevulinic acid (ALA), porphyrins) bound to metal nanoparticles. Plasmon resonance amplifies the excitation light (e.g., HeNe laser at 632.8 nm) at the nanoparticle surface, resulting in increased photoactivation of the PA drug molecules and improved therapy efficiency. This is intended to increase the relative efficacy or safety of therapy by requiring less radiation energy and intensity.[37], [52]

- **EPEP and Exciton-Plasmon Coupling:** Exciton-Plasmon Enhanced Phototherapy (EPEP) systems utilize the coupling of excitonic states with surface plasmons. This mechanism, potentially involving mixed plasmon-exciton states, can lead to efficient X-ray coupling for phototherapy.[38]

- **Plasmonic Nanobubbles (PNBs):** PNBs, generated by pulsed laser excitation of plasmonic nanoparticles, are used for diagnosis, removal, or mechanical damaging of tumors. They are studied for intracellular targeting and gene therapy, and for rapidly detecting and destroying drug-resistant tumors.[173]

## B. Drug Delivery and Functional Carriers

- **Drug Delivery Platforms:** Gold nanoparticles are described as a platform technology for tumor-targeted drug delivery. Drug molecules can be linked to the nanoparticles via linkers (e.g., biochemical bond, DNA bond, or antibody-antigen bond) that can be cleaved by photon radiation or by the internal chemical environment inside the cell (e.g., chemically labile bonds) for controlled drug release.[37], [52]

- **Stimuli-Responsive Magneto-Plasmonic Nanocarriers:** Hybrid structures like magneto-plasmonic nanocarriers (e.g., magnetic $Fe_3O_4$ core and gold/silver shell) are designed to be stimuli-responsive. They are utilized for multimodal image-guided therapy, including targeted drug delivery to organs like the brain, for treatments such as HIV.[174]

- **SERS/Magnetic Bifunctional Nanotubes:** Quasi-one-dimensional (1-D) plasmonic-magnetic (PM) nanotubes combine magnetic manipulation capabilities with enhanced optical sensing (SERS). These nanotubes are highly desirable for multiple-task applications in single-cell bioanalysis, magnetic manipulation and separation, and biosubstance or active agent delivery.[60]

**C. General Therapeutic Devices and Materials**

- **Plasmonic-Doped Biomaterials:** Materials like silk doped with plasmonic nanoparticles can be used in biomedical devices (e.g., Implantable Medical Devices (IMDs)). These materials can serve as light-activated heating elements for photothermal therapy and are relevant for monitoring and treating physiological conditions.[44]

- **Targeted Sterilization:** The plasmonics-enhanced mechanism can be applied to induce sterilization in joint regions of utensils by enhancing the production of internal UV light.[38]

## 3.6.3 Integrated Cell Analysis and Physiological Monitoring

Plasmonic systems provide non-invasive, real-time access to the complex dynamics of living cells and tissues.

- **Live Excitable Cell Monitoring:** Novel plasmonic devices, typically using a gold sensor on a prism within a fluidic module (microfluidic device), allow for the extraction, correlation, and quantification of action potentials in live excitable cells (e.g.,cardiac cells/cardiomyocytes (CMs). [115]

  ◦ **Ion Channel Activity:** The high sensitivity of plasmons to charge distribution (within 200 nm of the sensor surface) makes them suitable for sensing biologically driven charge variations, allowing the tracking of ionic activities of non-voltage gated and voltage-gated ion channels.[115]

  ◦ **Cell Health Analysis:** Properties essential for overall cell health analysis, including conductivity, excitability, rhythmicity, and contractility (e.g., beating in cardiac cells), can be extracted. Machine learning can be used to analyze this data to understand cardiovascular physiology and channelopathies.[115]

  ◦ **Drug Screening:** This method can be used to monitor the binding of drugs like blebbistatin, verapamil, and caffeine to CMs.[115]

- **Cell-Surface Interaction and Extracellular Sensing:**

  ◦ **Real-Time Secretion Imaging:** Nanoplasmonic sensors (LSPRi) are used for the quantitative imaging of protein secretions from single cells in real time. This technique measures absolute concentration and concentration gradient of secretions (e.g., cytokines) and is useful in developmental biology and identifying polarized secretions relevant to cell migration.[112]

  ◦ **Cell Manipulation and Sorting:** Plasmonic substrates are used in hybrid devices for on-chip concentration, manipulation, sorting, and sensing of particles (such as biological cells) by combining plasmonic heating (electrothermoplasmonics) with sensing.[62]

- **Bio-Actuation and Stimulation:**

  ◦ **Neural and Cardiac Stimulation:** Plasmonic technology is a breakthrough technology for stimulating electrically excitable biological cells. Potential applications include cochlear stimulation for cochlear implants, cardiac stimulation for cardiac pacemakers, muscle and nerve testing (e.g., plasmomyography), and retinal stimulation for retinal implants.[175]

  ◦ **Near-Infrared (NIR) Stimulation:** Plasmonic nanostructures (e.g., gold nanorods) can enhance infrared neural stimulation using localized surface plasmon resonance (LSPR).[175]

288-598-148-263